\documentclass[useAMS,usenatbib,fleqn]{mnras}
\usepackage{amsopn, natbib} \usepackage[utf8]{inputenc}
\usepackage{amsmath, bm} \usepackage{caption} \usepackage{amsfonts}
\usepackage{amssymb} \usepackage{graphicx} \usepackage{hyperref}
\usepackage{subfig} \usepackage{psfrag} \usepackage{array}
\usepackage{xcolor}
\usepackage{array, makecell}
\usepackage{lipsum} 

\usepackage[compact]{titlesec}  
\titlespacing{\section}{0pt}{8pt}{3pt}
\titlespacing{\subsection}{0pt}{6pt}{2pt}

\def\visataum{v^{\mathrm{f}} \left( \U_a, \tau_m \right)}
\def\avgvisai{\bar{\mathcal{V}} \left( \U_a, \nu_n  \right)}

\def\nvisait{\mathcal{N}^{t}(\U_a, \nu_n)}
\def\visai{\mathcal{M}\left( \mathbfit{U}_a, \nu_n \right)}
\def\sigvisai{\mathcal{S}\left( \mathbfit{U}_a, \nu_n \right)}
\def\fgvisai{\mathcal{F}\left( \mathbfit{U}_a, \nu_n \right)}
\def\visait{\mathcal{V}^{t}\left( \mathbfit{U}_a, \nu_n \right)}
\def\vis{\mathcal{V}\left( \mathbfit{U}, \nu \right)}

\def\v2{V_2\left(\mathbfit{U},\Delta \nu = 0\right)}

\def\dnu{\Delta \nu}
\def\Nc{N_{\mathrm{c}}}
\def\Ns{N_{\mathrm{s}}}
\def\dBdTnu{Q_{\nu_n}}

\def\pb{A\left(\Delta \hat{\mathbfit{n}},\nu  \right)}

\def\ap{\tilde{a}(\U , \nu )}

\def\appan{\tilde{a}(\U_a - \U', \nu_n )}

\def\U{\mathbfit{U}} 
\def\HI{{\ion{H}{I}~}}
\def\n{\hat{\mathbfit{n}}}
 
\def\m{\hat{\mathbfit{m}}}
\def\dn{\Delta \mathbfit{n}}
\def\Rt{R_{\mathrm{t}}}
\def\R{R (k_{\perp},k_{\parallel})}
\def\cR{\mathcal{R}(\Rt, t_{\mathrm{obs}})}
\def\tobs{t_{\mathrm{obs}}}
\def\bw{B_{\mathrm{bw}}}

\newcommand{\ppk}{\mathbfit{k}_{\perp}} 
\newcommand{\kpar}{k_{\parallel}}
\newcommand{\ppka}{\mathbfit{k}_{\perp a}} 
 
\newcommand{\nvppka}{k_{\perp a}} 
\newcommand{\kparm}{k_{\parallel m}}

\def\eg{{\it e.g.}\,}
\def\ie{{\it i.e.}\,}

\title[Foreground Avoidance]{Simulated Predictions for HI at $z=3.35$ with the Ooty Wide 
Field Array (OWFA) - II : Foreground Avoidance}

\author[Chatterjee et al.]{Suman Chatterjee$^{1,2,5}$, Somnath
  Bharadwaj$^{1,2}$, Visweshwar Ram
  Marthi$^{3,4,5}$\\ $^{1}$Department of Physics,
  Indian Institute of Technology Kharagpur, Kharagpur - 721 302,
  India.\\ $^{2}$Centre for Theoretical Studies, Indian Institute of
  Technology Kharagpur, Kharagpur - 721 302, India. \\
  $^{3}$ Canadian Institute for Theoretical Astrophysics, 60 St. George Street, 
  Toronto, ON M5S 3H8, Canada. \\  
  $^{4}$Dunlap Institute for Astronomy \& Astrophysics, 50 St. George Street, 
  Toronto, ON M5S 3H4, Canada.\\
  $^{5}$National Centre for Radio Astrophysics, Tata Institute of
  Fundamental Research, Post Bag 3, Ganeshkhind, Pune - 411 007, India.}
\begin{document}
\date{\today}

\pagerange{\pageref{firstpage}--\pageref{lastpage}} \pubyear{2018}

\maketitle
\label{firstpage}
\begin{abstract}
Considering the upcoming OWFA, we use simulations of the foregrounds and the $z = 3.35$ \HI 21-cm intensity mapping  signal  to identify the  $(k_{\perp},k_{\parallel})$ modes where the expected 21-cm power spectrum $P(k_{\perp},k_{\parallel})$  is substantially larger than the predicted foreground contribution. Only these uncontaminated $k$-modes are used for measuring $P(k_{\perp},k_{\parallel})$ 
 in the ``Foreground Avoidance"  technique. 
Though the foregrounds are largely localised within a wedge. we find that  the small leakage beyond the wedge surpasses  the 21-cm signal across a significant part of the $(k_{\perp},k_{\parallel})$ plane. 
The extent of foreground leakage is  extremely sensitive to the frequency window function used to 
estimate $P(k_{\perp},k_{\parallel})$. It is possible to reduce the leakage by making the window function narrower, however this comes at the expense of  losing a larger fraction  of the 21-cm signal. It is necessary to balance these competing effects to identify an optimal window function.  Considering a broad class of cosine window functions, we identify a six term window function as optimal for 21-cm power spectrum estimation with OWFA. 
 Considering only the $k$-modes where the expected 21-cm power spectrum exceeds the predicted foregrounds by a factor of $100$ or larger,  a $5\,\sigma$ detection of the binned power spectrum  is possible in the $k$ ranges $0.18 \leq k \leq 0.3 \, {\rm Mpc}^{-1}$  and $0.18 \le  k \le 0.8 \, {\rm Mpc}^{-1}$ with $1,000-2,000$ hours and $10^4$  hours  of observation  respectively.   
\end{abstract}


\section{INTRODUCTION} \label{sec:intro}

Intensity mapping with the neutral hydrogen (\HI) 21-cm radiation is a
promising tool to study the large scale structures in the
post-reionization Universe \citep{Bharadwaj2001b}. It holds the 
potential of measuring the Baryon Acoustic
Oscillation (BAO) that is imprinted in the \HI 21-cm power spectrum, 
and the comoving scale of BAO can be used as a standard ruler to 
constrain the evolution of the equation of state of dark energy \citep
{Wyithe2008, Chang2008, Seo2010, Masui2010}. Further a measurement of 
just the \HI 21-cm power spectrum can also be used to constrain 
cosmological parameters \citep{Bharadwaj2009,Visbal2009}. Higher order 
statistics such as the bispectrum holds the prospect of quantifying the
non-Gaussianities in the \HI 21-cm signal \citep{Ali2005, Hazra2012}.  
Using the \HI signal in cross-correlation with the WiggleZ galaxy survey data, 
the Green Bank Telescope (GBT) has made the first detection of the \HI 
signal in emission at $z \approx 0.8$ \citep{Chang2010, Masui2013}. \citet{Switzer2013} 
have constrained the auto-power spectrum of the redshifted \HI 21-cm radiation from 
redshift $z \sim 0.8$ with GBT.

The Giant Meterwave Radio Telescope (GMRT; \citealt{Swarup1991}) is
sensitive to the cosmological \HI signal from a range of redshifts in
the post-reionization era  \citep{Bharadwaj2003, Bharadwaj2005} 
and \citep{Ghosh2011a, Ghosh2011b} have carried out preliminary
observations towards detecting this signal from $z = 1.32$.
The Canadian Hydrogen Intensity Mapping Experiment (CHIME, 
\citealt{Newburgh2014} ,\citealt{Bandura2014})  aims to measure the BAO in the
redshift range $0.8 - 2.5$.   The future Tianlai \citep{Chen2012, Chen2015},
SKA1-MID  \citep{Bull2015ApJ}, HIRAX\citep{Newburgh2016} and MeerKLASS 
\citep{Santos2017}  also aim to measure the redshifted \HI
21-cm signal from the post-reionization era.  In this paper we consider the upcoming 
Ooty Wide Field Array (OWFA, \citealt{Subrahmanya2016a}) which aims to  measure the \HI signal 
from $z = 3.35$ .

The Ooty Radio telescope (ORT; \citealt{Swarup1971})   is a $530 \, {\rm m}$ long (North-South) and $30 \, {\rm m}$ wide (East-West) 
offset-parabolic cylinder which  is located on a hill whose slope roughly matches the latitude of the station ($11^{\circ}$).
Effectively, the axis of the cylinder is parallel to the  earth's rotation axis, making it equatorially mounted. The telescope is mechanically steerable in the East-West direction  by rotating the cylinder about its axis, allowing continuous tracking of a source on the sky.  Throughout this paper we consider observations which track a single field on the sky. This  allows the signal from different time instances to be coherently added to increase the signal to noise ratio (SNR).

The ORT is currently being upgraded to function as an interferometric array the OWFA. This upgrade   will result in two concurrent modes namely OWFA PI and PII. OWFA PI will be a 
linear array of $N_{\mathrm{A}} =  40$ antennas each with a rectangular aperture  $b \times d$, where 
$b = 30  \, {\rm m}$ and $d = 11.5 \, {\rm m}$.  The entire analysis of this paper is 
restricted to  OWFA PII which has  a larger number of antennas $N_{\mathrm{A}} = 264$, 
with smaller aperture  ($b = 30 \, {\rm m}$ and $d = 
1.92 \, {\rm m}$) arranged  with a spacing $d$ along the 
 North-South axis of the cylinder. The telescope operates at a nominal frequency of 
$\nu_{\mathrm{c}} = 326.5 \, {\rm MHz}$. We consider  $\Nc=312$ frequency channels  each 
of width $\Delta \nu=0.125 \, {\rm MHz}$ spanning a bandwidth of $\bw = 39 \, {\rm  MHz}$.

The details
of the antenna and hardware configuration can be found in
\citet{Prasad2011}, \citet{Subrahmanya2016a} and \citet{Subrahmanya2016b}.
Theoretical estimates \citep{Bharadwaj2015} 
predict that it should be possible to measure the amplitude of the
21-cm power spectrum  with $150 \, {\rm hours}$ of observations using
OWFA PII. A more recent study \citep{Sarkar2016a} indicates possible measurement
of the 21-cm power spectrum in
several different $k$ bins in the range $0.05 - 0.3 \, {\rm Mpc}^{-1}$
with $1,000 \, {\rm hours}$ of observations. \citet{Sarkar2018} have shown that
the cross-correlation of the redshifted HI 21-cm signal with OWFA PII
with the Lyman-$\alpha$ forest is detectable in a $200 \,{\rm hours}$-integration each in 
$25$ independent fields-of-view (FoV).

The complex visibilities are the primary quantities measured by any
radio-interferometric array like OWFA. It is possible to directly 
estimate the \HI 21-cm power spectrum from the measured visibilities 
\citep{Bharadwaj2001a, Bharadwaj2005}. \citet{Sarkar2016b} have proposed and implemented 
a new technique to estimate the OWFA \HI signal visibilities. Galactic and extragalactic foregrounds pose a 
severe challenge to the \HI 21-cm signal detection \citep{Ali2008, Ghosh2011b}. 
The theoretical estimates \citep{Ali2014} predict that the visibilities
measured at OWFA will be dominated by astrophysical foregrounds which
are expected to be several orders of magnitude larger than the \HI
signal. The astrophysical foregrounds are all expected to have a smooth frequency
dependence in contrast to the \HI signal. With the increasing 
frequency separation , the \HI signal is expected to decorrelate 
much faster than the foregrounds 
\citep{Bharadwaj2003}, a feature on which most foreground removal
techniques rely to distinguish between the foregrounds and the
HI signal. Modelling foreground spectra is challenging and is further complicated by the 
chromatic response of the telescope primary beam. \citet{Marthi2016a} (from now Paper I) have introduced 
a Multi-frequency Angular Power Spectrum (MAPS) estimator and demonstrated its ability, 
using an emulator (PROWESS; \citealt{Marthi2016b}), to accurately characterize the 
foregrounds for OWFA PI.

Several studies have  shown that the foreground contributions are expected to 
be largely confined within a wedge shaped region in the $(k_{\perp},k_{\parallel})$ plane
\citep{Datta2010,Vedantham2012,Morales2012, Parsons2012b, Trott2012}.  
In this work we focus on a conservative strategy referred to as ``foreground avoidance".
In this strategy only the $k$-modes where the predicted foreground contamination  is substantially below the expected 21-cm signal are used for power spectrum estimation. Ideally, one hopes to use the entire set of $k$-modes outside  the foreground wedge for estimating the 21-cm power spectrum. However, there are several factors which cause foreground leakage beyond the foreground wedge.   
 The chrormaticity   of the various foreground components and also the individual antenna elements causes foreground leakage beyond the wedge. 
The exact extent of this wedge is  still debatable 
(see \citealt{Pober2014} for a detailed discussion). The large OWFA FoV  makes it crucial  to address the wide-field effects for the foreground  predictions for OWFA. 
 On a similar note, the Fourier transform along the frequency axis used to calculate the cylindrical 
power spectrum introduces artefacts due to the  discontinuity in the measured visibilities at the edge of the band. It is possible to avoid this problem by introducing  a frequency window function which smoothly falls
to zero at the edges of the band. This issue has been studied by \citet{Vedantham2012} and \citet{Thyagarajan2013} who have proposed the Blackman-Nuttall (BN; \citealt{Nuttall1981}) window function. 
While the additional  frequency window does successfully mitigate the artefacts, it also 
introduces additional chromaticity which also contributes to  foreground leakage beyond the wedge boundary.

In this paper we have used simulations of  the foregrounds and  the \HI 21-cm signal  
expected for OWFA PII to quantify the extent  of the foreground  contamination outside the foreground wedge. The aim is to identify the $(k_{\perp}, k_{\parallel})$ modes which can be used for measuring the 21-cm power spectrum, and to  
 asses the prospects  of measuring  the 21-cm power spectrum using the foreground avoidance technique. 
Our all sky foreground simulations (described in Section~\ref{sec:fg})  incorporate  the  two most dominant components namely  the diffuse Galactic synchrotron emission and the extragalactic point sources. 
 This work improves upon the earlier work (Paper I) by introducing 
an all-sky foreground model. The simulated foreground  visibilities (described in Section~\ref{sec:vis})
incorporate the chromatic behaviour of both the sources and also the instrument.
The actual OWFA primary  beam pattern is unknown. We have carried out the entire study here using  two different  models for the primary beam pattern, we expect the actual OWFA beam pattern  to be in 
between the two different scenarios considered here. We have used the ``Simplified Analysis"  of  \citet{Sarkar2016b}
to simulate the \HI signal contribution to the visibilities (also described in Section~\ref{sec:vis}).
To estimate the 21-cm power spectrum from the the OWFA visibilities, in  Section~\ref{sec:3DPS} 
we  introduce and also validate a visibility based  estimator which  has been  constructed so as to 
eliminate the noise bias and provide an unbiased estimate of the 3D power spectrum. 

Our results (Section~\ref{sec:result}) show that the foreground leakage outside the wedge is extremely sensitive to the form of the frequency window function used for estimating the 21-cm power spectrum.  While the leakage can be reduced by making the window function narrower, this is at the expense of increasing the loss in the 21-cm signal. It is necessary to balance these two competing  effects in order to choose the optimal window function. In this paper we consider a broad class of  cosine  window functions each with a different number of terms. We introduce a figure of merit which allows us to quantitatively compare the performance of different window functions, and we use this to determine the optimal window function to estimate the 21-cm power spectrum using  OWFA. 
Considering the optimal window function, we finally quantify the prospects of measuring the 21-cm power spectrum using OWFA. The results are discussed and summarized in Section~\ref{sec:summary}.

We use the fitting formula of \citet{Eisenstein1999} for the
$\Lambda$CDM transfer function to generate the initial, linear
matter power spectrum. The  cosmological parameter values used are as
given in \citet{Planck2014}:$\Omega_m = 0.318$, $\Omega_b \, h^2 = 0.022$,
 $\Omega_{\lambda} = 0.682$, $n_s = 0.961$, $\sigma_8 = 0.834$, $h = 0.67$.

\begin{table*}
\begin{center}
{\renewcommand{\arraystretch}{1.5}%
 \begin{tabular}{||l | l ||  l | l ||} 
 \hline
 Symbol & Definition & Symbol & Definition\\
 \hline 
 \hline 
$b \times d$ & Aperture dimensions \, \, 30 m$\times$1.92 m & $N_{\mathrm{A}}$ & Number of antennas \, \, \, 264\\
\hline
$\nu_{\mathrm{c}}$ & Central frequency \, 326.5 MHz & $\U_a $ & Different baselines where  $ a=1,2,\ldots, N_{\mathrm{A}} - 1 $\\ 
\hline
$B_{\mathrm{bw}}$ & Bandwidth \, 39 MHz & $\Nc$ & Number of channels \, \, 312\\
\hline
$\dnu$ & Channel width \, \, 0.125 MHz &$\nu_n$ & Different frequency channels where $n=0,1,...,\Nc - 1$ \\
\hline
$N_r$ & Number of realization of simulations &$N_p$ & Number of terms in  frequency window function\\
\hline
  $\Ns$ & Number of time stamps & \makecell[l]{$r$ \\ $r'$} & \makecell[l]{Comoving distance to redshift 3.35 \, \, 6.84 Gpc \\ $\vert dr/d\nu \vert_{\nu = \nu_{\mathrm{c}}}$ \, \,  11.5 Mpc MHz$^{-1}$}  \\
\hline
$\tobs$ & Observation time & $\Delta t$ & Integration time\\
\hline
$P_{\mathrm{T}}(k_{\perp},k_{\parallel})$ & \HI 21-cm power spectrum & $P_{\mathrm{N}}(k_{\perp},k_{\parallel})$ & Noise power spectrum \\
\hline

$P_{\mathrm{L}}(k_{\perp},\kpar)$ & \makecell{Foreground leakage power spectrum} & $\R$ & $P_{\mathrm{T}}(k_{\perp},k_{\parallel})/ P_{\mathrm{L}}(k_{\perp},\kpar)$\\
\hline
$\Rt$ & Threshold value of $\R$ & $\cR$ & Ratio of SNR and SNR for MS6 window \\
\hline
 \hline 
 \end{tabular}}
 \end{center}
 \caption{Definitions for some  of the symbols used here.}
\label{tab:sym}
\end{table*}

\section{SIMULATIONS}\label{sec:fg}
The radiation from different astrophysical sources other than 
the redshifted cosmological \HI 21-cm radiation are collectively 
referred to as foregrounds. The most dominant contributions to the 
foregrounds at $326.5 \, {\rm MHz}$, come from the diffuse synchrotron 
from our own galaxy (Diffuse Galactic Synchrotron Emission; DGSE) 
and the extragalactic radio sources (Extragalactic Point Sources; EPS). 
The free-free emission from our galaxy  and from external galaxies 
are also larger than the \HI 21-cm signal. We exclude accounting the 
free-free emissions as a separate component in our analysis since they 
have power-law spectra similar to the other components  \citep{Kogut1996}. 
They are easily subsumed by the uncertainty in the discrete continuum source 
contribution and they make relatively smaller contributions to the foregrounds. 

\subsection{The Diffuse Galactic Synchrotron Emission }\label{subsec:dgse}
The diffuse galactic synchrotron emission (DGSE) arises from the 
energetic charged particles (produced mostly by supernova explosions) 
accelerating in the galactic magnetic field \citep{Ginzburg1969}. 
Various observations at $150 \, {\rm MHz}$ \citep{Bernardi2009, Ghosh2012, 
Iacobelli2013b, Choudhuri2017} have quantified $\mathcal{C}_{\ell}$  
the angular power spectrum of brightness temperature fluctuations of the DGSE. 
Based on these we have modelled the DGSE using 
\begin{equation}
\mathcal{C}_{\ell}(\nu)  = 513 \, {\rm mK}^2 \,  \left(\frac{1000}{\ell}\right)^{2.34} \, \left(\frac{150 \, {\rm MHz}}{\nu}\right)^{5.04} \,  ,
\label{eq:maps1}
\end{equation}
where the amplitude and the $\ell$ power law index are from \citet{Ghosh2012},  
whereas for the frequency spectral index we have used the results from \citet{Rogers2008}.  
Various studies indicate that the amplitude and slope have different values in different 
patches of the sky (\eg \citealt{LaPorta2008}, \citealt{Choudhuri2017}), and so also the 
spectral index \citep{DeOliveiraCosta2008}. These variations will introduce additional 
angular and frequency structures. However, in our simulations we have used fixed values 
across the entire sky. 

\begin{figure*}
\psfrag{ell}{$\ell$}
\psfrag{clinmksq}{$\mathcal{C}_{\ell} \, \, {\rm mK}^2$}
\psfrag{FFmodel}{Model}
\psfrag{FFFsimulation}{Simulation}
\includegraphics[scale=0.4]{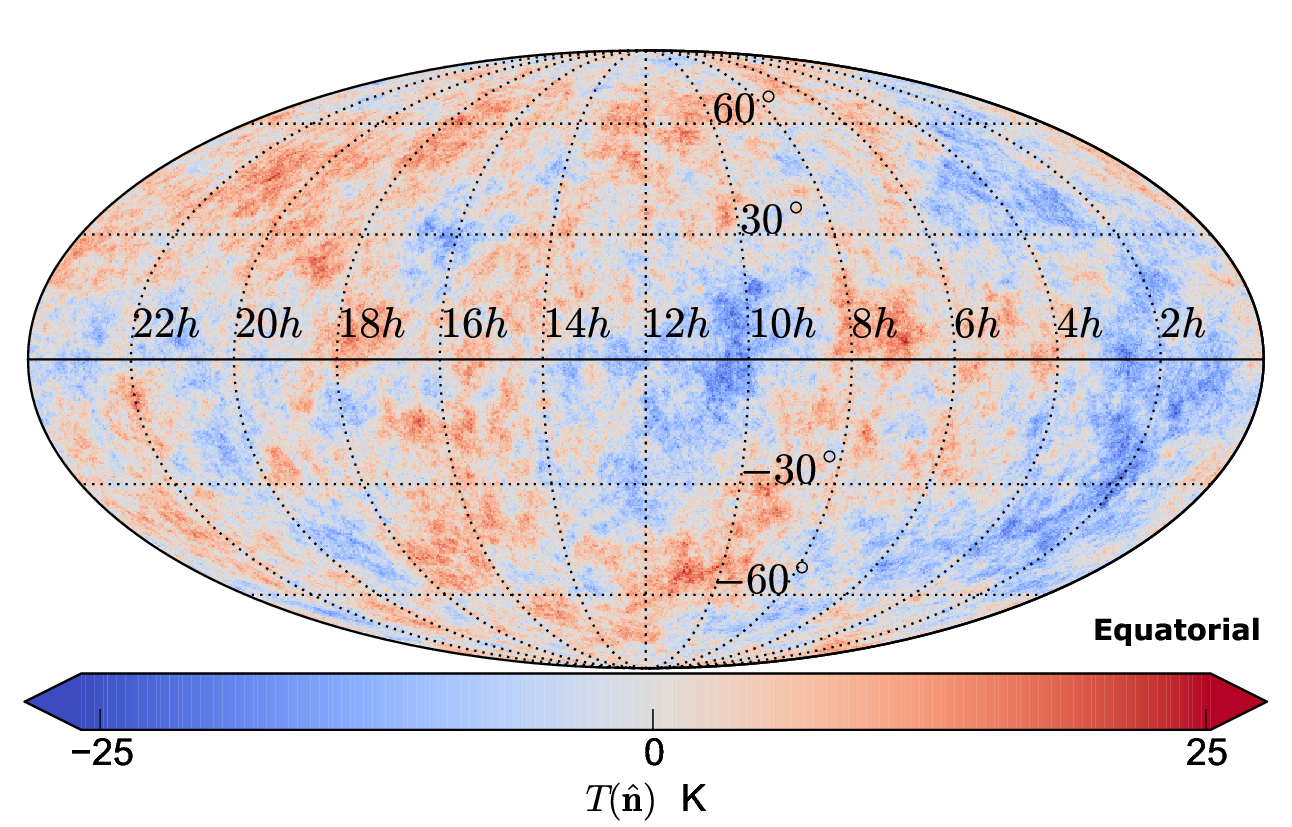}
\includegraphics[scale=0.7]{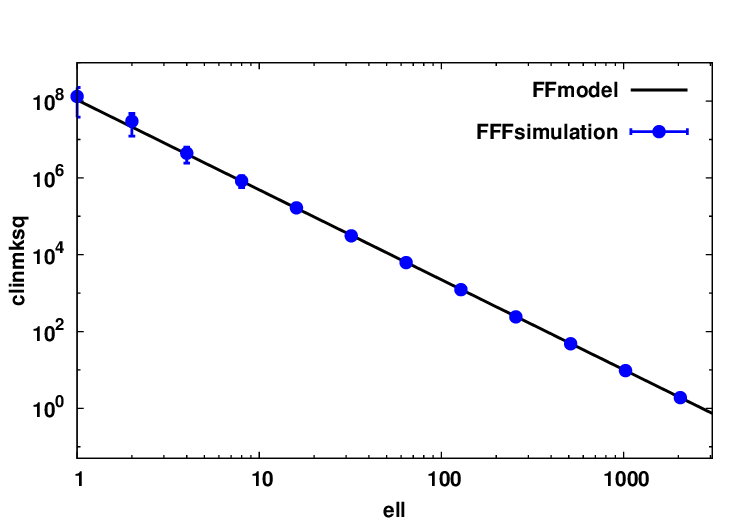}
\caption{The left panel shows a single realization of the simulated DGSE map for the 
nominal frequency of $\nu_{\mathrm{c}} = 326.5 \, {\rm MHz}$. In the right panel, the line
shows the angular power spectrum $\mathcal{C}_{\ell}$ of the input DGSE model 
(eq.~\ref{eq:maps1}) at $\nu_{\mathrm{c}}$ while the points with the error-bars are the 
mean and standard deviation obtained from the simulations.}
\label{fig:dgse}
\end{figure*}

We simulate the DGSE using the package Hierarchical Equal Area isoLatitude Pixelization of a
sphere \citep[HEALPix;][]{Gorski2005}, where we represent the entire sky using $12582912$ 
pixels of size $3.435'$. We assume that the brightness temperature  fluctuations of the DGSE 
are a Gaussian Random Field and used the SYNFAST routine of HEALPix to generate different 
statistically independent realizations of the brightness temperature fluctuations at 
$\nu_{\mathrm{c}}$. These were scaled to obtain the brightness temperature fluctuations at 
the other frequency channels in the observing bandwidth of OWFA. The left panel of 
Figure~\ref{fig:dgse} shows a particular realization of the simulated DGSE maps and the 
right panel shows a comparison of $\mathcal{C}_{\ell}$ values estimated from the simulations 
(in points) and the input model (in solid line) at $\nu_{\mathrm{c}}$. We use $20$ 
statistically independent realizations of the DGSE simulations to estimate the 
mean values and $1\sigma$ error bars shown here.

\subsection{Extragalactic Point Sources}
The extragalactic point sources (EPS) are expected to dominate the $326.5 \, {\rm MHz}$ 
sky at most of the angular scales of our interest. These sources are a mix of 
normal galaxies, radio galaxies, quasars, star-forming galaxies, and other objects, which 
are unresolved by the OWFA. We model the differential source count $dN/dS$ of the sources 
using the fitting formula given by \citet{Ali2014},
\begin{eqnarray} 
  \frac{dN}{dS}= \left \{\begin{array}{ll}
   4000(\frac{S}{1{\rm Jy}})^{-1.64} ({\rm Jy} \cdot{\rm Sr})^{-1} & 
   3 \, {\rm mJy} \le \,S \le 3 \, {\rm Jy}  \\ 
   134(\frac{S}{1 {\rm Jy}})^{-2.24} ({\rm Jy} \cdot{\rm Sr})^{-1}  &   
    {10 \, \mu \rm Jy} \le S \le 3  {\rm mJy} ,
         \\
    \end{array}\right. 
  \label{eq:sc}
\end{eqnarray}
where they fit  the $325 \,{\rm MHz}$ differential source counts 
measured by \citet{Sirothia2009}. This is consistent with the WENSS $327 \, 
{\rm MHz}$ differential source count, (Figure 9 of \citealt{Rubart2013}).  For the 
sources  below $ 3 \,\rm mJy$, they fit the $1.4 \, {\rm GHz}$ source counts 
from extremely deep VLA observations (\citealt{Biggs2006}) and extrapolate it 
to $326.5 \, {\rm MHz}$. Here we assume that the sources with flux $S >  
3 \, {\rm mJy}$ make the major contribution to foregrounds, and only consider 
sources with $S > 3 \, {\rm mJy}$. We assume that the spectral nature of such 
sources can be modelled (spectral behaviour) as a power law $S_{\nu} \propto 
\nu^{\alpha}$, where for each source we randomly assign a value of $\alpha$ 
drawn from a Gaussian distribution with mean $-2.7$ and ${\it r.m.s.} = 0.2$ 
\citep{Olivari2018}. The angular clustering of radio sources at low flux densities 
is not well known. To make an estimate, we use the angular correlation function $w(\theta)$ 
measured from NVSS, which can be approximated as $w(\theta)\approx (1.0 \pm 0.2) 
\times 10^{-3} \, \theta^{-0.8}$ \citep{Overzier2003}, for which the angular 
power spectrum  $w_{\ell}$ has been calculated to be  $w_{\ell} \approx 1.8 \times 10^{-4} \ell^{-1.2}$ 
\citep{Blake2004b, Olivari2018}.
The EPS contribution to the brightness temperature fluctuations can be decomposed 
into two parts, namely (a) the Poisson fluctuations due to the discrete 
nature of the sources, and (b) a fluctuation due to the angular clustering of the 
sources. The simulations were carried out using HEALPix with the same specifications 
as mentioned in Section~\ref{subsec:dgse}. Based on the differential source counts 
(eq.~\ref{eq:sc}) we expect $3145728$ sources in the sky map, corresponding to  a 
mean $0.25$ sources per pixel. We have simulated $100$ times the expected number 
of sources (mean $25$ per pixel), and assigned them flux values which are randomly  
drawn from the differential source count  (eq.~\ref{eq:sc}) and whose spectral index 
values are assigned randomly as discussed earlier.  To incorporate the angular 
clustering  we generate realizations of Gaussian random fluctuations $\delta_p$ 
($p$ labels the  pixels) corresponding to the angular power spectrum $w_{\ell}$. 
We distribute the simulated sources among the pixels by assigning  $25 (1+ \delta_p)$ 
sources from the simulated source list to pixel $p$,  and so on covering all 
the pixels in the sky map.  Finally, we have randomly selected $3145728$ sources 
from the simulated source list. The  brightness temperature distribution resulting 
from these simulated sources now incorporates  both the Poisson fluctuation and the 
angular clustering of the sources.

\begin{figure}
\centering
\psfrag{ell}{$\ell$}
\psfrag{clinmksq}{$C_{\ell} \, \, {\rm mK}^2$}
\psfrag{clcomparison}{.}
\psfrag{total}{Total}
\psfrag{diffuse}{DGSE}
\psfrag{poisson}{Poisson}
\psfrag{cluster}{EPS}
\psfrag{Scut3}{$ $}
\psfrag{Scut2}{$S_c = 0.1\,{\rm Jy}$}
\psfrag{Scut1}{$S_c = 0.01\,{\rm Jy}$}

\psfrag{oeps}{EPS}
\psfrag{simulationforDiffuseFFF}{Simulation(DGSE)}
\psfrag{AnalyticforDiffEps}{Analytic(Total)}
\psfrag{epstot}{Total}

\includegraphics[scale=0.86]{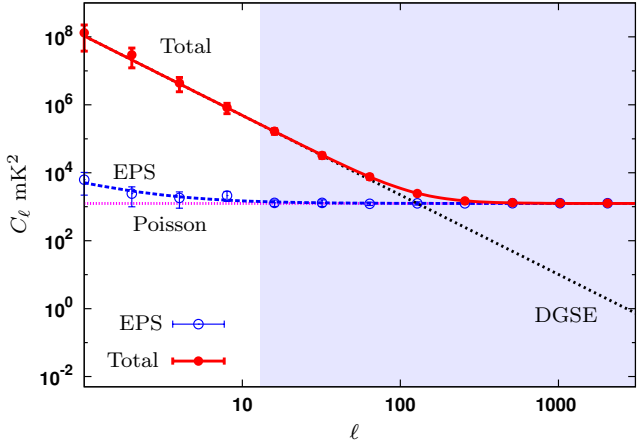}
\caption{This shows the angular power spectrum $\mathcal{C}_{\ell}$ of the brightness 
temperature fluctuations of the foreground components. The hollow and filled circles 
show the mean EPS and the total foreground (\ie EPS+DGSE) contributions to 
$\mathcal{C}_{\ell}$ with $1\sigma$ error bars estimated from $20$ statistically 
independent realizations of the simulations, the analytical predictions are shown in 
different line-styles as indicated in the figure. The shaded region bounds the $\ell$ 
range probed by OWFA PII.}
\label{fig:eps2}
\end{figure}
Figure ~\ref{fig:eps2} shows the mean $\mathcal{C}_{\ell}(\nu_{\mathrm{c}})$ 
along with   $1\sigma$ error bars  estimated from $20$ statistically  independent 
realizations of the foregrounds. This is compared with the theoretical predictions 
for the different components, as well as the predictions for the total expected 
$\mathcal{C}_{\ell}$. We see that the DGSE dominates at large angular scales \ie  
$\ell \,< 600$,  while the EPS dominates at small angular scales \ie $\ell \, \ge  600$.


\section{VISIBILITY SIMULATION}\label{sec:vis}
OWFA is a linear  array of $N_{\mathrm{A}}$ antennas arranged without any intervening gap along the length  of a North-South parabolic cylinder. Each antenna has a rectangular aperture of dimension $b \times d \, = \, 30 \, 
{\rm m} \times 1.92 \, {\rm m}$ illuminated by four end  to end linear dipoles (Figure 4 of Paper I). The spacing between the centers of two adjacent antennas also is $d$, and  we  have the smallest baselines 
 $\U_1 = ({d}/{\lambda})\, \hat{\mathbfit{z}}$ 
 where $\lambda$ is the observing wavelength and we adopt a Cartesian coordinate system which is tied to the telescope with the the unit vectors $\hat{\mathbfit{z}}$  and $\hat{\mathbfit{y}}$ respectively along the length and breadth of the cylinder.  The unit vector  $\hat{\mathbfit{x}}$ points  perpendicular to the antenna aperture, this direction   lies along the celestial equator.

The baselines $\U_a= a \times \U_1$ measured at  OWFA are all multiples of $\U_1$ with 
 $ a=1,2,..., N_{\mathrm{A}} - 1 $. The OWFA baselines  have a high degree of  redundancy   \ie  we have 
 $(N_{\mathrm{A}} -a)$ different  antenna pairs which correspond to the same 
 baseline $\U_a$.  The OWFA visibilities $\visait$ are measured at $n=0,1,...,\Nc - 1$ different frequency channels each with a respective central frequency  $\nu_n$.  Following  Paper I, we express the measured visibilities as 
\begin{equation}
\visait  =  \visai + \nvisait
\label{eq:mvis1}
\end{equation}
where $\visai$ refers to model visibilities originating from the sky signal, 
and $\nvisait$ is the additive system  noise contribution. Here the label $t = 0, 1, 2, \ldots , \Ns-1$ in $\visait$  
denotes distinct measurements of the visibilities each corresponding to a different time stamp 
 and $\Ns$ denotes the total number of time stamps. We note that there are several effects like calibration errors,  ionospheric fluctuations and man made RFI which maye also contribute to the measured visibilities in eq.~(\ref{eq:mvis1}), however we do not consider these here. 
We also consider that a fixed field is being tracked throughout the observation.
The model visibility which originates from the sky signal is given by \citep{Perley1989}
\begin{equation}
\visai = \dBdTnu \int d\Omega_{\n} \, 
  T\left(\n,\nu_n \right) A\left(\dn,\nu_n
\right) e^{-2 \pi i \U_a\cdot \dn },
\label{eq:vis1}
\end{equation}
where, $\dBdTnu = 2 k_{\mathrm{B}} / \lambda_n^2$ is the conversion factor from
brightness temperature to specific intensity in the Rayleigh - Jeans
limit, $ T\left(\hat{\mathbfit{n}},\nu_n \right)$ is the brightness
temperature distribution on the sky along the direction of the unit vector $\n$ which has  sky coordinates ${\rm (RA,DEC)}=(\alpha,\delta)$,
$d\Omega_{\n}$ is the elemental solid angle in the direction $\n$,  
and  $\dn = \n - \m$ where  $\m$ is the unit vector in the pointing direction of the antennas which also corresponds to    $x$. Throughout this work we assume that $\m$ points towards the position ${\rm (RA,DEC)}=(0,0)$ on the sky.

The  model visibilities $\visai$  can further be considered 
to be the sum of two parts 
\begin{equation}
\visai = \fgvisai \, +\,  \sigvisai  \, .
\label{eq:modvis1}
\end{equation}
which refer to the the foreground and the  \HI signal respectively.

\begin{figure*}
\psfrag{mb2}{$-d/2$}
\psfrag{b2}{$d/2$}
\psfrag{xx}{$z$}
\psfrag{yy}{$\,$}
\psfrag{illumination pattern}{$\! \!\! \! \! \! \! \! \!$Illumination pattern}
\psfrag{pfv}{$\! \!  0.5$}
\psfrag{one}{$\! \! 1.0$}
\psfrag{mpib4}{$-45^{\circ}$}
\psfrag{pib4}{$+45^{\circ}$}
\psfrag{mpib2}{$-90^{\circ}$}
\psfrag{pib2}{$+90^{\circ}$}
\psfrag{AindB}{$ \! \! \! \! \! \! \! \! \! \! \! \! A(\Delta n, \nu_{\mathrm{c}})$~(dB)}
\psfrag{theta}{$\quad \delta $}
\psfrag{primary beam pattern}{$\! \!\! \! \! \! \! \! \!$Primary beam pattern}
\psfrag{Uniform}{$\! \!\! \! \! \! \! \! \!$ Uniform}
\psfrag{triangle}{$\! \!\! \! \! \! \! \! \!\! \! \! \! \! \! $ Triangular}
\psfrag{0}{$0$}
\psfrag{mtf}{$ \! \! \! \!  -25$}
\psfrag{mfif}{$\! \! \! \!  -50$}
\centering
\includegraphics[scale=0.85, angle = 0]{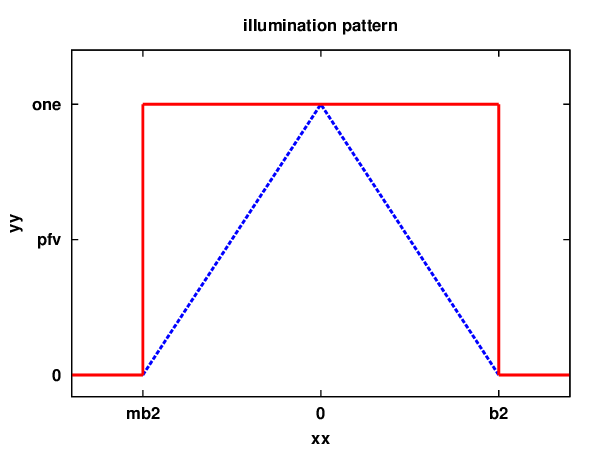}
\hskip 4mm 
\includegraphics[scale=0.85, angle = 0]{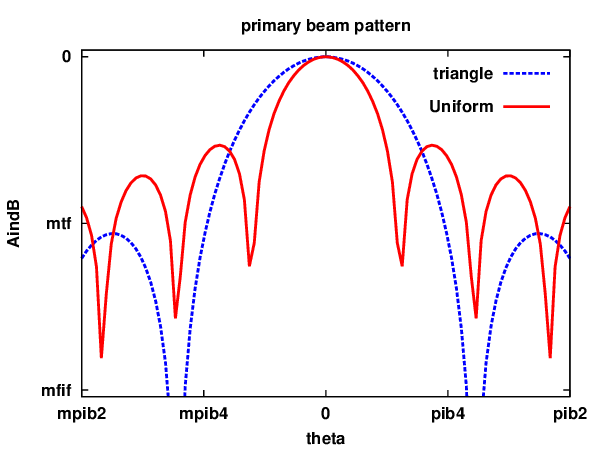}
\caption{The left panel shows the Uniform (red solid line) and Triangular 
(blue dashed line) illumination patterns considered here.   
The right panel shows the corresponding primary beam patterns (Table~\ref{tab:pb})  along the North-South direction.}
\label{fig:Pbeam}
\end{figure*}

The foreground contribution $\fgvisai $  is highly sensitive to the 
telescope's primary beam pattern  \citep{Berger2016}.  The actual OWFA primary beam pattern $A\left(\dn,\nu_n \right)$ is currently unknown, and we have considered   two different possibilities for the predictions presented here. The first model for $A\left(\dn,\nu_n \right)$ (Table~\ref{tab:pb}) is based on the simplest assumption that  the OWFA antenna  aperture is uniformly  illuminated by the dipole feeds, which results in the ``Uniform" sinc-squared primary beam pattern   considered in several earlier works (\citealt{Ali2014}, Paper I,\citealt{Chatterjee2018a}). 
In reality, the actual illumination  pattern is expected to fall away from the aperture centre resulting in a wider field of view as compared to the Uniform illumination. In order to assess how this affects the foreground predictions and  foreground mitigation, we have considered  
a ``Triangular" illumination pattern (Figure~\ref{fig:Pbeam} ) for which we have a broader 
sinc-power-four  primary beam pattern (Table~\ref{tab:pb}).

\begin{table*}
\begin{center}
{\renewcommand{\arraystretch}{1.2}%
 \begin{tabular}{||c | c | c ||} 
 \hline
  Illumination & Uniform & Triangular \\ 
\hline
Primary Beam pattern $A(\dn,\nu) = $ & ${\rm sinc}^2 \left( \pi b \, \Delta n_y / \lambda \right) 
{\rm sinc}^2 \left( \pi d\,\Delta n_z /\lambda \right)$  & ${\rm sinc}^4 \left( \pi b \,
\Delta n_y / 2 \lambda  \right) {\rm sinc}^4 \left( \pi d\,   \Delta n_z / 2 \lambda \right)$ \\

\hline

  Aperture  power pattern $\ap = $ & $\left( \lambda^2/b d \right) \,  \Lambda \left( U_y \lambda/b \right) 
\Lambda \left( U_z \lambda/ d \right)$ & $\left( 64 \lambda ^2/ b d \right) \,  G \left( U_y \lambda /b \right)
 G \left( U_z \lambda / d \right)$\\

\cline{1-3}

 &$ \Lambda(x) = 
\begin{cases}
1-\vert x \vert & {\rm for}  \, \vert x \vert < 1 \\
0  & {\rm for} \,  \vert x \vert \geq 1 \\
\end{cases} $ & $ G(x) = 
\begin{cases}
1/6 - \vert x \vert^2 + \vert x \vert^3 & {\rm for}  \, \vert x \vert < 1/2 \\
1/3 -\vert x \vert +\vert x \vert^2 - \vert x \vert^3/3 & {\rm for}  \, \vert x \vert \geq 1/2 \\
0  & {\rm for} \,  \vert x \vert \geq 1 \\
\end{cases} $\\
\hline
FWHM,  $\eta$,  $\tilde{\eta}$ & $1.55^{\circ} \times 24^{\circ}$  , \, $1$, 
\, $32.49$ & $2.25^{\circ} \times 35^{\circ}$ , \, $9/16$, \,  $19.86$\\
\hline
\end{tabular}}
\end{center}
\caption{Here $\Delta n_y, U_y$  and $\Delta n_z, U_z$  are 
respectively the $y$ and $z$ components of $\dn$ and $\U$.
FWHM is the full width at half-maximum of $A(\dn,\nu)$. 
$\eta$ (eq.~\ref{eq:noise}) and $\tilde{\eta}$ (eq.~\ref{eq:apn-np}) are the aperture efficiency 
and a dimensionless factor respectively.}
\label{tab:pb}
\end{table*}

Considering both the Uniform and the Triangular beam patterns, Figure~\ref{fig:Pbeam} shows the variation 
of $A(\dn, \nu)$ with $\delta$ (\ie along  the North-South direction)
for fixed $\alpha=0$ and $\nu=\nu_{\mathrm{c}}$. Comparing the two beam patterns we find that 
the Uniform main lobe subtends  $\sim \pm 24^{\circ}$ (FWHM) whereas this is approximately double 
 $\sim \pm 35^{\circ}$ (FWHM) for Triangular. The number of side lobes is also found to decrease from Uniform to Triangular.  The Uniform and Triangular beam patterns represent two extreme cases, and the  actual OWFA beam pattern will possible be somewhere in between these two extreme cases both in terms of the extent of the main lobe and the number of side lobes.

We use the simulated foreground maps (Section~\ref{sec:fg}) to compute 
$\fgvisai$ using 
\begin{eqnarray}
\fgvisai &=& \dBdTnu \, \Delta \Omega_{\mathrm{pix}} \, \sum _{p} \, T(\alpha_p,
\delta_p, \nu_n) \nonumber \\
&\times& A(\alpha_p, \delta_p, \nu_n) e^{-2 \pi i U_a \, (\sin
  \delta_p )}\, ,
  \label{eq:vis_sim1}
\end{eqnarray}
where $\Delta \Omega_{\mathrm{pix}}$ is the solid angle subtended by
each simulation pixel. The  pixels in the simulated maps are labeled using $p$ with corresponding 
 ${\rm (RA,DEC)}_p=(\alpha_p,\delta_p)$ and the sum is over all the pixels in the simulation.

Considering $\sigvisai$, the \HI signal contribution to the model visibilities, we have simulated these using the flat-sky approximation (FSA). An earlier work \citep{Chatterjee2018a} has carried out a full spherical harmonic analysis for OWFA to find that the differences from the FSA are at most within $10 \%$ at the few smallest baselines and they are much smaller at the other larger  baselines. Using 
 $\dn=\bm{\theta}$ which is now a 2D vector on the plane of the sky, eq.~(\ref{eq:vis1}) reads 
\begin{equation}
 \sigvisai= Q_{\nu_n} \, \int \, d^2\bm{\theta} \, T\left(\bm{\theta},\nu_n \right) A\left(\bm{\theta},\nu_n \right) 
e^{-2 \pi i \, \U_a \cdot \bm{\theta} } \, .
\label{eq:sec1-vis2}
\end{equation}
whereby  $\sigvisai$  is the  Fourier transform of $[Q_{\nu_n} \, T\left(\bm{\theta},\nu_n \right) \, A\left(\bm{\theta},\nu_n \right)]$. We can express this as 
 a convolution \citep{Ali2014},
\begin{equation}
\sigvisai = Q_{\nu_n} \int d^2 \U' \, \appan \, \tilde{T}(\U', \nu_n) \, ,
\label{eq:vis3}
\end{equation} 
where $\tilde{T}(\U', \nu)$ is now the Fourier transform of $T\left(\bm{\theta},\nu \right)$,  
and the aperture power pattern $\tilde{a}(\U, \nu) =\int d^2\bm{\theta} \, e^{- 2 \pi i \U \cdot \bm{\theta}} \, A\left(\bm{\theta},\nu \right)$ (Table~\ref{tab:pb}).

In a recent work \citet{Sarkar2016b} have proposed an analytic technique to simulate $ \sigvisai$  the 
\HI signal contribution to the visibilities which is based on the FSA. Here we have used the  ``Simplified Analysis" 
presented  in Section 2 of \citet{Sarkar2016b}.  This uses the 
eigenvalues and the eigenvectors of the predicted two-visibility correlation matrix 
$S_2(\U_a,\nu_n,\nu_{n'})=\langle  \sigvisai \mathcal{S}^{*}(\U_a,\nu_{n'}) \rangle$
to simulate multiple statistically independent realizations of  $ \sigvisai$.  
The Simplified Analysis used here ignores  the correlation between the \HI signal at adjacent baselines
and also the non-ergodic nature of the \HI visibility signal along the frequency axis, both of these have however 
been included in the ``Generalized Analysis" presented in \citet{Sarkar2016b}.  We note that it is necessary to  diagonalize  the entire covariance matrix between the visibilities at all the baselines and frequency channels in order to incorporate the correlations between the \HI signal at the adjacent baselines.  This is computationally intensive and we have avoided this by adopting  the Simplified Analysis which considers each baseline separately significantly reducing the dimension of the covariance matrix.

The two-visibility correlation $S_2(\U_a,\nu_n,\nu_{n'})$ is related to the 21-cm brightness temperature power 
spectrum  $P_{\mathrm{T}}(\mathbfit{k})$ \citep{Bharadwaj2001a,Bharadwaj2005}. For OWFA we have \citep{Ali2014}, 

\begin{eqnarray}
S_2(\U_a,\nu_n,\nu_{n'}) &=& Q^2_{\nu_{\mathrm{c}}} \int \frac{d^3k}{(2\pi)^3} \, \vert \tilde{a}(\U_a - 
\frac{\ppk r}{2\pi }, \nu_{\mathrm{c}}) \vert^2 \nonumber \\
&\times& P_{\mathrm{T}}(\ppk,\kpar) \, e^{i r' \kpar (\nu_{n'} - \nu_n)}   \,  
\label{eq:viscor}
\end{eqnarray}
Here $\ppk$ which is 
the component of a 3D wave vector $\mathbfit{k}$ perpendicular to the line of sight, 
can be associated with the baselines $\U$ available at OWFA as $\ppk =  2 \pi \U/r$ , 
where $r = 6.84 \, {\rm Gpc}$ is the comoving distance to $z = 3.35$ and $r' = 
\vert d r / d \nu \vert_{\nu = \nu_{\mathrm{c}}} = 11.5 \, {\rm Mpc \, MHz}^{-1} $ 
sets the conversion scale from the frequency separation to comoving distance in the radial direction.
Here $\kpar$ is the line of sight component of a 3D wave vector $\mathbfit{k}$.

The \HI 21-cm brightness temperature power spectrum $P_{\mathrm{T}}(k_{\perp},k_{\parallel})$ 
is modelled as \citep{Ali2014} 	
\begin{equation}	
P_{\mathrm{T}}(k_{\perp},k_{\parallel})=  \bar{T}^{2} b^{2}_{\HI} \bar{x}_{\HI}^{2} [1+\beta \mu^{2}]^{2} P(k) \, , 
\label{eq:HI_PS}
\end{equation}
where $\mu = \kpar/k$, $\bar{T}=4.0 mk (1+z)^{2} \left(\frac{\Omega_{b} h^{2}}{0.02}\right) 
\left(\frac{0.7}{h}\right) \left(\frac{H_{0}}{H(z)}\right)$, 
$b_{\HI}= 2$ is the linear bias, $\bar{x}_{\HI}= 2.02 \times 10^{-2}$ is 
the mean neutral hydrogen fraction and $P(k)$ is the power spectrum of the underlying dark matter density 
distribution. The term $\left( 1+\beta \mu^2 \right)$ arises due to of the effect of HI peculiar 
velocities,  and $\beta = f(\Omega)/b_{\HI}$
is the linear redshift distortion parameter, where $f(\Omega)$ is the 
dimensionless linear growth rate. We use
$\beta=0.493$ and $f(\Omega) = 0.986$ throughout this paper. 
It is worth mentioning that, detailed simulations \citep{Castorina2017, DSarkar2018, Villaescusa_Navarro2018, Modi2019} predict that the non-linear effects due the redshift 
space distortion becomes significant at high-$\kpar$ modes. These non-linearities and the shot 
noise present at high-$k$ modes are not included in the 21-cm power spectrum 
model considered here (Eq.~\ref{eq:HI_PS}). However we expect those high-$k$ modes 
$(k > 1.0\,{\rm Mpc}^{-1})$ to be noise dominated and their contribution to the SNR is 
expected to be small.

The noise contribution  $\nvisait$ in each visibility is assumed to be  an independent 
complex Gaussian random variable with zero mean. The real part (or equivalently the imaginary part) 
of the noise contribution has a {\it r.m.s.} fluctuation,
\begin{equation}
\sigma_{\mathrm{N}}(\U_a) = \frac{\sqrt{2} \, k_{\mathrm{B}} \, T_{\mathrm{sys}}}{\eta \, A \, \sqrt{\dnu \, \Delta t \, (N_{\mathrm{A}} - a)}}
\label{eq:noise}
\end{equation}
where $T_{\mathrm{sys}}$ is the total system temperature, $k_{\mathrm{B}}$ is the Boltzmann constant, $A = b \times d$ 
is the physical collecting area of each antenna, $\eta $ is the aperture efficiency (Table~\ref{tab:pb}) 
with $\lambda^2/\eta A=\int A\left(\bm{\theta},\nu \right) d^2 \theta$ and $\Delta t = 16 \, {\rm s}$ is 
the correlator integration 
time. The OWFA baselines are highly redundant \citep{Ali2014, Subrahmanya2016b} and 
the factor $1/\sqrt{(N_{\mathrm{A}} - a)}$ in $\sigma_{\mathrm{N}}(\U_a)$ accounts for the redundancy in the baseline distribution. 
We expect $T_{\mathrm{sys}}$ to have a value around $150\,{\rm K}$, and we use this value for the estimates 
presented here.

\section{3D POWER SPECRTUM ESTIMATION}\label{sec:3DPS}
We now discuss how the measured visibilities $\visait$ are used 
to estimate the 3D power spectrum $P(\ppka, \kparm)$.
Considering a particular baseline $\U_a$ and frequency $\nu_n$, the  different time-stamps 
$\visait$  contain the same sky signal, only the system noise 
is different. We first average over the different time-stamps to reduce the data volume 
\begin{equation}
\avgvisai = \frac{1}{\Ns} \sum_{t=0}^{\Ns-1} 
\visait \, ,
\end{equation} 
here $\Ns$ denotes the total number of time stamps.
The  visibilities $\avgvisai$ are then Fourier transformed along the frequency axis to   
obtain the visibilities $\visataum$ in delay space
 \citep{Morales2004}
\begin{equation}
\visataum = \left( \dnu \right) \sum_{n = 0}^{\Nc - 1} e^{2 \pi i \tau_m \nu_n} F(\nu_n) \avgvisai \, .
\label{eq:vis5}
\end{equation}
where the delay variable $\tau_m$  takes values 
$\tau_m = m/\bw$ with $-\Nc/2 < m \leq \Nc/2$. The Fourier transform here assumes that the visibility signal is periodic in frequency with a period equal to the bandwidth $\bw$. The measured visibilities  $\avgvisai$, however, do not satisfy this requirement. This introduces a discontinuity in the values of the visibilities and also their derivatives 
at the edge of the frequency band. As noted in several earlier works \citep{Vedantham2012, Thyagarajan2013}, 
these discontinuities  introduce artifacts which  result   in  foreground leakage outside the foreground wedge. In addition to this, several other features like the frequency dependence of both the foreground sources and  the telescope's primary beam pattern also contribute to the foreground  leakage. However in this paper we entirely  focus on the leakage arising from the discontinuities at the boundary of the   frequency band. 
The leakage from these discontinuities can be reduced  \citep{Vedantham2012} by introducing  $F(\nu)$ (eq.~\ref{eq:vis5}) which is  a frequency window function that  smoothly falls to zero at 
the edges of the band making the product $[F(\nu_n) \, \avgvisai]$ effectively 
continuous at the edge of  the band. 
Earlier works \citep{Vedantham2012, Thyagarajan2013} 
show that  the Blackman-Nuttall (BN) \citep{Nuttall1981} 
window function is a promising candidate for  power spectrum estimation, and this   is expected to reduce 
the foreground leakage by $7-8$ orders of magnitude.  However,  discontinuities in the various derivatives  persist and  as we shall see later, the  BN window function fails to reduce  the foreground leakage to a level below the \HI signal 
expected at  OWFA.  In order to investigate if this problem can be overcome by considering other window functions, 
we have considered a broader set of cosine window functions 
\begin{equation}
F(\nu_n)= \sum_{p=0}^{N_p-1}  (-1)^p \, A_{p} \, \cos\big(\frac{2 n p \pi}{\Nc-1}\big) \, ,
\label{eq:win}
\end{equation}
each having different coefficients $A_p$ and number of terms $N_p$. Since the band is divided into  an even number ($N_c$) of frequency channels, the channel with index $N_\mathrm{c}/2$ is considered as the centre frequency where the window function peaks.  Note that this is equivalent to DFT-even \citep[see \eg][]{Harris1978}.
Here we have  considered the  Blackman-Harris  4-term  window function (BH4), and a family of Minimum Sidelobe (MS) window functions   (MS5, MS6 and MS7). Of these, the  BN \citep{Paul2016} and BH4 \citep{Eastwood2019} have been used  extensively in recent observational studies.  
 Table \ref{tab:window} shows the  coefficients \citep{Albrecht2001} of these window functions considered here.  

\begin{table*}
\begin{center}
{\renewcommand{\arraystretch}{1.5}%
 \begin{tabular}{||c | c c c c c||} 
 \hline
 Coefficient & \multicolumn{5}{c||}{ The cosine frequency window functions $F(\nu_n)$} \\
 \hline
 $A_p$ & BN & BH4 & MS5 & MS6 & MS7 \\ 
   & $(N_p = 4)$ & $(N_p = 4)$ & $(N_p = 5)$ & $(N_p = 6)$ & $(N_p = 7)$  \\
 \hline\hline
 $A_0$ & $3.6358\times 10^{-1}$ & $3.5875\times 10^{-1}$ & $3.2321\times 10^{-1}$ & $2.9355\times 10^{-1}$ & $2.7122\times 10^{-1}$ \\
 \hline
 $A_1$ & $4.8918\times 10^{-1}$ & $4.8829\times 10^{-1}$ & $4.7149\times 10^{-1}$ & $4.5193\times 10^{-1}$ & $4.3344\times 10^{-1}$\\ 
 \hline
 $A_2$ & $1.3659\times 10^{-1}$ & $1.4128\times 10^{-1}$ & $1.7553\times 10^{-1}$ & $2.0141\times 10^{-1}$ & $2.1800\times 10^{-1}$ \\
 \hline
 $A_3$ & $1.0641\times 10^{-2}$ & $1.1680\times 10^{-2}$ & $2.8496\times 10^{-2}$ & $4.7926\times 10^{-2}$ & $6.5785\times 10^{-2}$ \\
 \hline
 $A_4$ &        &            & $1.2613 \times 10^{-3}$ & $5.0261\times 10^{-3}$ & $1.07618 \times 10^{-2}$\\
 \hline
 $A_5$ &        &            &          & $1.3755\times 10^{-4}$ & $7.7001 \times 10^{-4}$ \\
 \hline
 $A_6$ &        &            &          &            & $1.3680 \times 10^{-5}$ \\ [1ex] 
 \hline
\end{tabular}}
\end{center}
\caption{The coefficients of the different window functions used in this work.}
\label{tab:window}
\end{table*}
\begin{figure*}
\psfrag{BN4}{\kern-3pt \small BN}
\psfrag{BH4}{\kern-3pt \small BH4}
\psfrag{BH5}{\kern-3pt \small MS5}
\psfrag{BH6}{\kern-3pt \small MS6}
\psfrag{BH7}{\kern-3pt \small MS7} 
\psfrag{mth}{$-5$}
\psfrag{th}{$5$}
\psfrag{mth2}{$-10$}
\psfrag{th2}{$10$}
\psfrag{wtaubB}{$\vert \tilde{f}(\tau_m)\vert~({\rm dB})$}
\psfrag{tt}{\kern-15pt \kern-10pt Delay Channel $(m)$}
\psfrag{pfv}{$\! \!  0.5$}
\psfrag{one}{$\! \! 1.0$}
\psfrag{mbwb2}{$0$}
\psfrag{bwb2}{$312$}
\psfrag{mbwb4}{$78$}
\psfrag{bwb4}{$234$}
\psfrag{nu}{$156$}
\psfrag{wfqn}{$ \! \! \! \! \! \! \! F(\nu_n)$}
\psfrag{frequency}{\kern-5pt \kern-10pt Frequency Channel $(n)$}
\psfrag{Uniform}{$\! \!\! \! \! \! \! \! \!$ Uniform}
\psfrag{triangle}{$\! \!\! \! \! \! \! \! \!$ Triangle}
\psfrag{0}{$0$}
\psfrag{mhun}{$ \! \! \! \!  -100$}
\psfrag{mfif}{$\! \! \! \!  -50$}
\psfrag{filter-fqn}{\kern-3pt \kern-5pt Frequency space}
\psfrag{filter-tau}{\kern-3pt \kern-3pt Delay space}
\includegraphics[scale=0.9, trim = 0.5cm 0cm 0.5cm 0cm, clip=true]{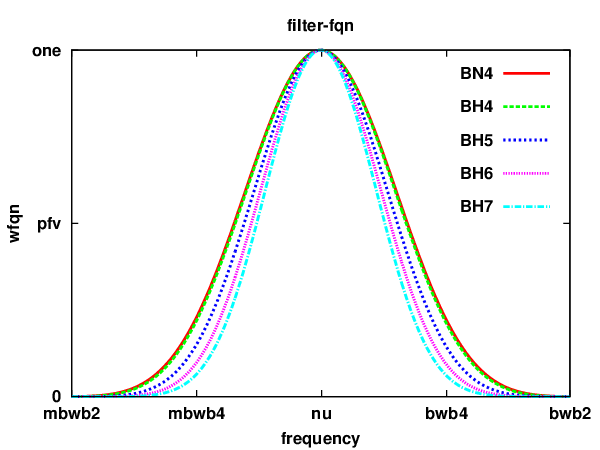}
\includegraphics[scale=0.9, trim = 0.0cm 0cm 0.5cm 0cm, clip=true]{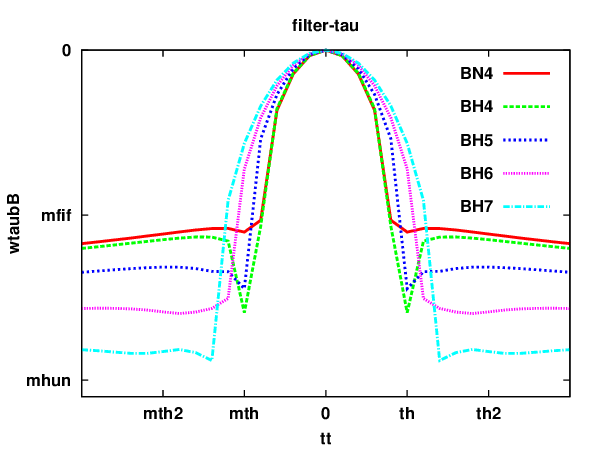}
\caption{The left panel shows the window functions $F(\nu_n)$ (mentioned in the legend)
as a function of channel number $n$ and the right panel shows the $\tilde{f}(\tau)$ for a small number of delay 
channels. $\tilde{f}(\tau)$ is normalized to unity at the central delay channel.}
\label{fig:w_func}
\end{figure*}

The left panel of Figure \ref{fig:w_func} shows 
the different window functions $F(\nu)$ considered here. As discussed earlier, we 
see that the window function smoothly goes down to zero towards the edge of the band.
An immediate consequence of introducing the window function $F(\nu)$ is the loss of signal, primarily 
towards the edge of the frequency band. Considering the window functions $F(\nu)$ in the order 
shown in Table \ref{tab:window}, we see that the  $F(\nu)$ gets successively narrower as 
we move from BN to MS7. We expect the suppression at the edge of the band to be more 
effective as the window function gets narrower, however this comes at an expanse of increasing 
sensitivity loss.

Considering the delay space visibilities $v\left( \U_a, \tau_m \right)$ without the window 
function (\ie $F(\nu) = 1$ in eq.~\ref{eq:vis5}), we have \citep{Choudhuri2016b}
\begin{equation}
\visataum = \frac{1}{\bw} \, \sum_{m' = -\Nc/2} ^{\Nc/2 - 1} \tilde{f}\left( \tau_m - \tau_{m'} \right) \, 
v\left( \U_a, \tau_{m'} \right) \, .
\label{eq:conv}
\end{equation}
We see that $\visataum$ is related to $v\left( \U_a, \tau_m \right)$ through a convolution 
with $\tilde{f}(\tau_m)$ which is the Fourier transform of the frequency window $F(\nu)$.
This convolution smoothens  out the signal over the  width of  $ \tilde{f}( \tau_m)$.
 Considering the \HI signal, the delay space visibilities $v\left( \U_a, \tau_m \right)$ and $v\left( \U_a, \tau_{m'} \right)$
at two different delay channels $\tau_m$ and $\tau_{m'}$  are predicted  to be uncorrelated (\eg \citealt{Choudhuri2016b}). The convolution in  eq.~(\ref{eq:conv}) however introduces correlations in $\visataum$ at two different values of the delay channel, the extent of this correlation is restricted within  the width of $ \tilde{f}( \tau_m)$. 
 The right-hand panel of Figure \ref{fig:w_func} 
shows the amplitude of $\tilde{f}(\tau_m)$ for the different window functions considered here. 
 We see that $\tilde{f}(\tau_m)$ peaks at $m= 0$,  and the values of $\tilde{f}(\tau_m)$ are very small beyond the primary lobe which is typically a few delay channels wide. 
This primary lobe of $\tilde{f}(\tau_m)$ gets successively wider as we move from BN to MS7
\ie the window function $F(\nu)$ gets successively narrower. The 
BN window function has the narrowest $\tilde{f}(\tau_m)$  and $\visataum$ will be correlated 
upto $m \approx \pm 4$, whereas this extends to $m \approx \pm 7$ for MS7 which is the widest in delay space. The finite width of $\tilde{f}(\tau_m)$ also leads to a loss of \HI signal at the smallest $\tau_m$ values which correspond to the largest frequency separations.  Figure \ref{fig:w_func}  illustrates the fact that $\tilde{f}(\tau_m)$ widens and  this loss in \HI signal increases as we move from the BN to the MS7 window function. 

The delay space visibility $\visataum$ is related to the \HI 21-cm brightness temperature fluctuation 
$\Delta T_b(\ppka,\kparm)$ where 
$\ppka =  2 \pi \mathbfit{U}_a/r$ and $\kparm = 2 \pi \tau_m/r'$ \citep{Morales2004}, and we can use this to estimate $P(\ppka, \kparm)$  the 3D power spectrum of the sky signal. 
Considering the auto-correlation of $\visataum$ we have 
\begin{equation} 
\langle \vert \visataum  \vert^2 \rangle = C_F^{-1} \, \left[  P(\ppka, \kparm) + P_{\mathrm{N}}(\ppka, \kparm)\right] \, ,
\label{eq:v2_same}
\end{equation}
with 
\begin{equation}
    C_F^{-1} = \frac { \dnu \, \, \sum_n \vert F(\nu_n)\vert^2 [Q_{\nu_{\mathrm{c}}}^2 \,\,
\int d^2 \mathbfit{U} \, \vert \tilde{a}\left( \mathbfit{U},\nu_c \right) \vert^2] }{r^2 \, r'}\, ,
    \label{eq:cf}
\end{equation}
and the noise power spectrum is 
\begin{equation}
P_{\mathrm{N}}(\ppka, \kparm)=    
C_F \, \left( \frac{\dnu}{\Ns} \right)^2 \,
\sum_{n=0}^{\Nc -1 } \sum_{t=0}^{\Ns -1} \, \langle \vert \nvisait \vert^2 \rangle \, 
\vert F(\nu_n) \vert^2 \,.
\label{eq:pn}
\end{equation}

 The angular brackets  $\langle \ldots \rangle$ here denote an ensemble average over different random realizations of the \HI 21-cm signal. We can use $ \vert \visataum  \vert^2$ to estimate the \HI 21-cm power spectrum $ P(\ppka, \kparm)$ except for  the term $P_{\mathrm{N}}(\ppka, \kparm)$  which arises  due to the system noise  (eq.~\ref{eq:mvis1}) in the measured visibilities. This introduces a positive noise bias which needs to be accounted for before we can use eq.~(\ref{eq:v2_same}) to estimate $ P(\ppka, \kparm)$.

We use eq.~(\ref{eq:v2_same}) to  define $\hat{P}\left( \U_a, \tau_m \right)$   the 3D power spectrum estimator as 
\begin{eqnarray}
& & \hat{P}\left( \U_a, \tau_m \right) = C_F \, [ \vert \visataum  \vert^2  \nonumber \\
&-& \left(\frac{\dnu}{\Ns} \right)^2 \, \,  
\sum_{n=0}^{\Nc -1 } \sum_{t=0}^{\Ns - 1} \, \vert \visait \vert^2 \vert F(\nu_n) \vert^2   ] \, . \quad 
\label{eq:p_same}
\end{eqnarray}
The second term in the right-hand side of eq.~(\ref{eq:p_same}) is introduce to exactly 
subtract out  the noise bias in eq.~(\ref{eq:v2_same}). The estimator $\hat{P}\left( \U_a, \tau_m \right)$  therefor 
provides an unbiased estimate of the power spectrum,    and we have
\begin{equation}
P(\ppka, \kparm) = \langle \hat{P}\left( \U_a, \tau_m \right) \rangle \,. 
\label{eq:pks}
\end{equation}
 In addition to the noise bias,  some signal  also is  subtract out, however the fraction of the total visibility correlation signal that is lost is of the order of $\sim 1/\Ns$  which is extremely small for a long observation. For example we have $\Ns \sim 10^5 $ for $\tobs=1,000$ hours of observation with an integration time of $\Delta t=16 \, {\rm s}$.  It is worth noting that the correlation between the adjacent baselines can also be used to obtain additional estimates of the power spectrum \citep{Ali2014},  however we have not considered this possibility here. Appendix \ref{sec:appendix} presents analytical expressions for the variance of the estimator,  this is useful for predicting the uncertainty in the estimated power spectrum.

\subsection{Validating the estimator}
\begin{figure}
\psfrag{V2V2UaUb}{$\tiny P(k) {\rm mK}^2 {\rm Mpc}^{3}$}
\psfrag{kk}{$\qquad \qquad \qquad k\,{\rm Mpc}^{-1}$}
\psfrag{sperically averaged} {\kern-10pt \kern-10pt \small Spherically averaged}
\psfrag{Binned}{\kern-15pt \kern-10pt \kern-5pt \small Binned model}
\psfrag{simulation}{\kern-10pt \small Simulation}
\psfrag{error}{$\Delta$}
\psfrag{Uniform}{Uniform}
\psfrag{Triangular}{Triangular}
\psfrag{tobs1}{\kern-5pt$\tobs = 1,000 \, {\rm hrs}$}
\psfrag{tobs2}{\kern-5pt$\tobs = 10,000 \, {\rm hrs}$}
\centering
 \includegraphics[scale=0.75, angle = 0,trim=0.15cm 0.1cm 0.0cm 0.2cm, clip=true]{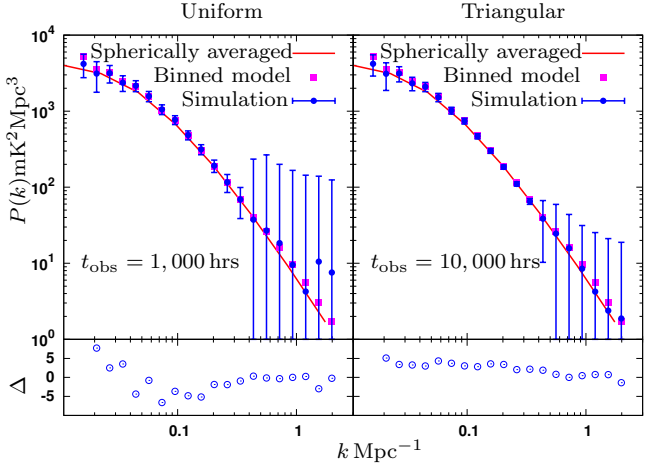}
\caption{Considering the power spectrum, the upper panels show a comparison of the spherically averaged input model, the  binned input model  and that estimated from the simulations.  $N_r=1,000$ statistically independent realizations of the simulation were used to estimate the mean and $1\sigma$ error bars shown here.  The points in the bottom panels show $\Delta$ (eq.~\ref{eq:err}), 
which quantifies the deviation between the binned input model and the simulated power spectrum.}
\label{fig:pk}
\end{figure}

To validate the \HI signal simulations and the 3D power spectrum estimator we have carried out simulations of the \HI signal visibilities using the prescription described in Section~\ref{sec:vis}  
considering both the Uniform and the Triangular  illuminations.   For both  cases we have simulated $N_r = 1,000$ statistically independent realizations of the \HI signal visibilities including the system noise component.
To reduce the data volume and the computation,  we have considered a total observation time of $\tobs=1,000$ hours  with an integration time of  $\Delta t=1\,{\rm hour}$ for the Uniform illumination,  and  $\tobs=10,000$ hours with $\Delta t=10\,{\rm hours}$ for the triangular illumination respectively. 
In both cases we have $\Ns=1,000$ which implies that we have $1/\Ns = 0.1 \%$ loss in the visibility correlation due to the term which cancels out the noise bias. As mentioned earlier, we expect this loss to be even smaller in actual observations where $\Ns$ will be much larger.  The upper panels of Figure~\ref{fig:pk} show the spherically averaged input model 21-cm brightness temperature power spectrum  $P_{\mathrm{T}}(k)$  as a function of $k$. The figure also shows the 
binned input model power spectrum where  we have considered  $P_{\mathrm{T}}(k)$ at the $(k_{\perp},k_{\parallel})$ modes corresponding to the OWFA baselines and delay channels, and binned these into $20$ equally spaced logarithmic bins. 

The simulated \HI  21-cm signal visbilities $\sigvisai$  considered here only contain the auto-correlation signal, as mentioned earlier the correlations between the adjacent baselines have not been incorporated here.  We have used the simulated visibilities in  eq.~(\ref{eq:p_same}) to estimate the power spectrum.  
The upper panels of Figure~\ref{fig:pk} show the binned power spectrum $P(k)$ estimated from the simulations, the left and the right panels show the results for Uniform and Triangular illuminations respectively.   The  $N_r$ realizations of the simulations were  used to estimate the mean and the $1\sigma$ error bars shown in the figure.   In both the cases we find that the estimated  power spectra are in good agreement with the input power spectrum. The error-bars  at the smallest $k$ bins are somewhat large due to the cosmic variance, though we see that a detection is possible here.  At large $k$ the errors exceed the expected power spectrum, and a detection is not possible within the $\tobs$ considered here. In both the illuminations  we see that  the errors are relatively small in the $k$ range $0.05-0.3 \, {\rm Mpc}^{-1}$ which is most favourable for measuring the power spectrum with OWFA \citep{Sarkar2016a}.  
The lower panels of Figure~\ref{fig:pk} show the dimensionless ratio 
\begin{equation}
\Delta = \frac{\delta P(k) \, \sqrt{N_r}}{\sigma} \, .
\label{eq:err}
\end{equation}
Here $N_r$ is the number of realizations of the simulations, $\delta P(k)$ is the difference between the estimated and the input model 
power spectrum. Ideally we expect this to have a spread of the order of $\sigma/\sqrt{N_r}$ around zero arising from statistical fluctuations. The normalized dimensionless ratio $\Delta$ is thus expected to have a variation  of order unity provided the estimator provides an unbiased estimate of the power spectrum. We find that the values of $\Delta$ in the lower panels are 
distributed within $\pm 5$ at all the bins except for that at the  smallest $k$ value. The power spectrum is possibly underestimated at the lowest few baselines because the estimator ignores the convolution with the aperture power pattern which is included in the visibility signal (see eq.~\ref{eq:viscor} and also \citealt{Choudhuri2014}). This deviation is however seen  to be well within the $1\sigma$  error-bars for $\tobs=1,000$ hours of observation (upper left panel of  \ref{fig:pk}). Overall we conclude that our simulations validate the power spectrum estimator presented here. 

\section{RESULTS} \label{sec:result}
\begin{figure*}
\psfrag{pkmksqmpcc}{$\tiny P(k_{\perp}, \kpar)\, {\rm mK}^2 \, {\rm Mpc}^{3}$}
\psfrag{V2V2UaUb}{$\tiny P(\nvppka, \kpar) {\rm mK}^2 {\rm Mpc}^{3}$}
\psfrag{kperp}{$k_{\perp} \, {\rm Mpc}^{-1}$}
\psfrag{kpar}{$\, \, \, \kpar \, {\rm Mpc}^{-1}$}
\psfrag{kpara}{$\, \, \, \, \kpar \, {\rm Mpc}^{-1}$}
\psfrag{mmmm0}{$m = 0$}
\psfrag{mmmm4}{$m = 4$}
\psfrag{mmmm14}{$m = 14$}
\psfrag{mmmm12}{$m = 12$}
\psfrag{base5type1}{ ${\scriptstyle k_{\perp} = 0.057}$} 
\psfrag{base10type2}{ ${\scriptstyle k_{\perp} = 0.115}$}
\psfrag{base35type3}{ ${\scriptstyle k_{\perp} = 0.402}$}
\psfrag{bl60PII}{ $a = 60$}
\psfrag{total}{Total}
\psfrag{HI}{\HI} 
\psfrag{DGSE}{DGSE}
\psfrag{EPS}{$\,$}
\psfrag{Sc3jy}{EPS}
\psfrag{Sc01jy}{$\!\! \!\! \!\! \!{\scriptstyle S_c = 100 \, {\rm mJy}}$}
\psfrag{Sc001jy}{$ \! \! {\scriptstyle S_c = 10 \, {\rm mJy}}$}
\psfrag{phaseI}{PI}
\centering
\includegraphics[scale=0.78, angle = 0]{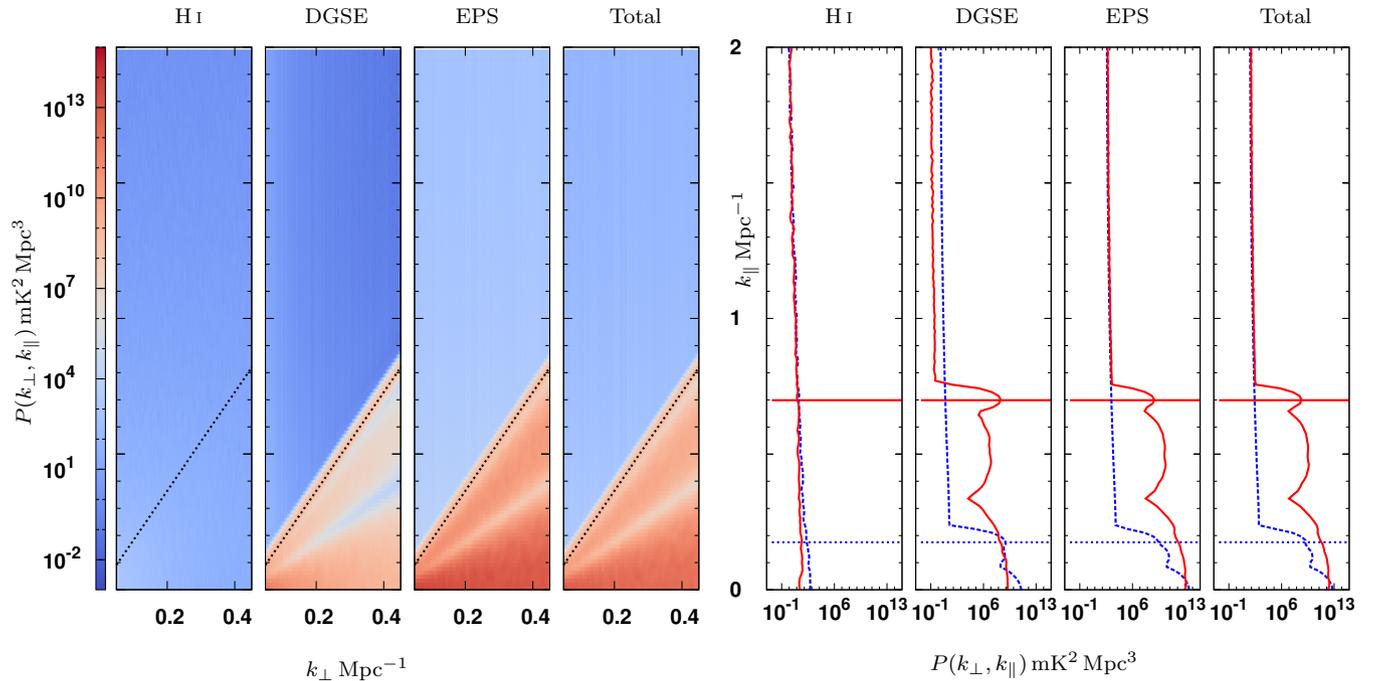}
\caption{Predictions of the component-wise contributions to the 3DPS 
$P(k_{\perp}, \kpar)$  from the \HI signal, the DGSE, the EPS and the total 3DPS. 
Panels in left show the cylindrical power spectrum 
$P(k_{\perp}, \kpar)$. The dotted lines mark the approximate wedge boundaries 
(eq.~\ref{eq:a1}). Panels in right show vertical sections through 
the left panels for fixed  $k_{\perp} = 0.095 \, {\rm Mpc}^{-1}$ (in dashed lines), 
$0.34 \, {\rm Mpc}^{-1}$ (in solid lines). 
The horizontal solid lines and dotted lines in the right 
panels indicate the approximate wedge boundaries (eq.~\ref{eq:a1}) for the above 
mentioned $k_{\perp}$ modes.}
\label{fig:FG_PI}
\end{figure*}
We first  focus on the \HI signal and foreground predictions,  and we have not included  the noise contribution  here.  Considering the Uniform illumination and the BN window function,  the left panel of Figure~\ref{fig:FG_PI}  shows the  predicted  cylindrical power spectrum $P(k_{\perp}, \kpar)$ averaged over $20$ statistically independent realizations of the simulations 
for the \HI signal, the 
individual  DGSE and  EPS foreground components  and the total sky signal. 
 We see that  the foregrounds are largely confined within the  ``Foreground Wedge" \citep{Datta2010}. 
 The foreground  contamination would  be restricted  to  $\kpar = 0$ if the foregrounds were spectrally flat in the absence of the instrument  {\it i.e.} the visibilities $\vis$ were independent of frequency. However, the fact that  
the baselines $U = d \, \nu / c$  change with frequency introduces  a frequency dependence in $\vis$ even if the sky signal is frequency independent. The  foreground simulations here include both   the  $\nu$ scaling of $U$ as well as the intrinsic $\nu$  dependence of the sky signal,  and as a consequence 
the foreground contribution to $P(k_{\perp}, \kpar)$ extends out along $k_{\parallel}$
onto a wedge which is expected to be  bounded by
\begin{equation}
k_{\parallel} = \left[ \frac{r \, \sin(\theta_l)}{r' \, \nu_{\mathrm{c}}} \right] 
\,k_{\perp} 
\label{eq:a1}
\end{equation}
in the $(k_{\parallel},k_{\perp})$ plane \citep{Datta2010, Vedantham2012, Morales2012,
Parsons2012b, Trott2012} where   $\theta_l$  refers to the largest angle (relative to the 
telescope's pointing direction) from which  we have a  significant foreground contamination. 
Here we consider $\theta_l = 90^{\circ}$ as the horizon limit. We see (Figure~\ref{fig:FG_PI}) 
that there is a very large foreground contribution at $\kpar = 0$, and the foregrounds beyond 
this are largely contained within a wedge. The dotted line in the figure shows the wedge 
boundary predicted by eq. (\ref{eq:a1}): we see that the boundary of the simulated foreground 
wedge is located  beyond the dotted line. The primary beam pattern $A(\dn, \nu )$, the 
intrinsic frequency dependence of the sources introduced through the spectral index $\alpha$ and 
$F(\nu)$ all introduce additional frequency dependence (or chromaticity) in $\vis$ 
which enhance  the extent of the foreground wedge beyond that predicted by eq.~(\ref{eq:a1}). 
We also notice that there are several structures visible inside the foreground wedge.

The right panels show vertical sections through the left panels \ie they show $P(k_{\perp}, 
\kpar)$ as a function of $\kpar$ for fixed $k_{\perp}$ values. We have chosen $k_{\perp} 
= 0.095$ and  $\,0.34 \, {\rm Mpc}^{-1}$ (dashed and solid lines respectively) for which the horizontal 
lines show the corresponding wedge boundaries predicted by eq.~(\ref{eq:a1}). Considering the foregrounds, 
 the $\kpar$ dependence of 
$P(k_{\perp}, \kpar)$ shows two peaks, the first at  
$\kpar = 0$ and the second at the wedge 
boundary. The second peak corresponds to what is known as 
the ``pitch fork" effect \citep{Thyagarajan2015, Thyagarajan2015b}, which is seen to be more  
prominent at  the larger baseline. The foreground 
wedge is found to extend by $\Delta \kpar \simeq 0.1 \, {\rm Mpc}^{-1}$ beyond the horizontal 
lines. In addition to this, we find oscillatory structures within the wedge where the 
$\kpar$ values of the dips  correspond to the nulls in the primary beam 
pattern (\ie replace $\theta_l$ in eq.~\ref{eq:a1} with $\theta_1$, $\theta_2 \ldots$ the 
angular positions of the various nulls of the primary beam pattern).  Considering large 
$\kpar$ beyond the wedge boundary, in all cases we find that  $P(k_{\perp}, \kpar)$ drops to a small value which 
does not change very much with $\kpar$. This small value of $P(k_{\perp}, \kpar)$ arises due to the 
foreground leakage beyond the wedge. For DGSE the value of  $P(k_{\perp}, \kpar)$ decreases 
with increasing $k_{\perp}$. This reflects  the fact that the DGSE contribution decreases with 
increasing $\ell$ $(\mathcal{C}_\ell \propto \ell^{- 2.34})$. In contrast, the EPS contribution, 
which is Poisson dominated,  does not change much with $k_{\perp}$.

Considering the \HI signal (Figure~\ref{fig:FG_PI}) 
we find that, the foreground contribution is $\sim 10^{10}$ times larger at $\kpar = 0$ 
and other points within the wedge boundary. We also find that  the foreground leakage  remains
$ \sim 10^{2}$ times larger than the \HI signal beyond the wedge boundary. This implies that the 
BN window is not a suitable choice for \HI  power spectrum detection with OWFA. 
  Although the BN window function ensures that the values of the filtered visibilities are continuous  at the boundary of the frequency band, discontinuities still persist in the various derivatives.  The leakage  in  $P(k_{\perp}, \kpar)$ seen at large  $\kpar$ beyond the wedge boundary
arises from these discontinuities.  Further,    Figure~\ref{fig:FG_PI} shows  an interesting feature that the leakage power $P(k_{\perp}, \kpar)$ appears to be proportional to $P(k_{\perp}, 0)$ which is the foreground power  at $\kpar=0$. We note that such a  behaviour is not surprising  as we expect the leakage amplitude to be proportional to the magnitude of the discontinuities which in turn are expected to be proportional to the overall foreground amplitude. Here $P(k_{\perp}, 0)$ provides a measure of the total foreground power (\ie amplitude squared, eqs. \ref{eq:vis5} and \ref{eq:p_same}).

In this paper we consider the possibility of controlling the discontinuities at the band edges  
 by suitably tailoring  the  window function. 
This leads us to investigate the  possibility 
of using higher term window functions for \HI  power spectrum detection with OWFA. To this end we 
make a comparative study of the expected foreground leakage for the set 
of window functions discussed earlier (eq.~\ref{eq:win} and Table~\ref{tab:window}). 
\begin{figure*}
\psfrag{pkmksqmpcc}{$\R$}
\psfrag{V2V2UaUb}{$\tiny P(\nvppka, \kpar) {\rm mK}^2 {\rm Mpc}^{3}$}
\psfrag{kperp}{$k_{\perp} \, {\rm Mpc}^{-1}$}
\psfrag{kpar}{$\, \, \, \kpar \, {\rm Mpc}^{-1}$}
\psfrag{kpara}{$\, \, \, \, \kpar \, {\rm Mpc}^{-1}$}
\psfrag{BH4}{BH4}
\psfrag{MS5}{MS5} 
\psfrag{MS6}{MS6}
\psfrag{MS7}{MS7}

\psfrag{PhaseI}{Uniform}
\psfrag{PhaseII}{Triangular}
\centering
\includegraphics[scale=0.75, angle = 0, trim = 0.5cm 0cm 0cm 0cm, clip=false]{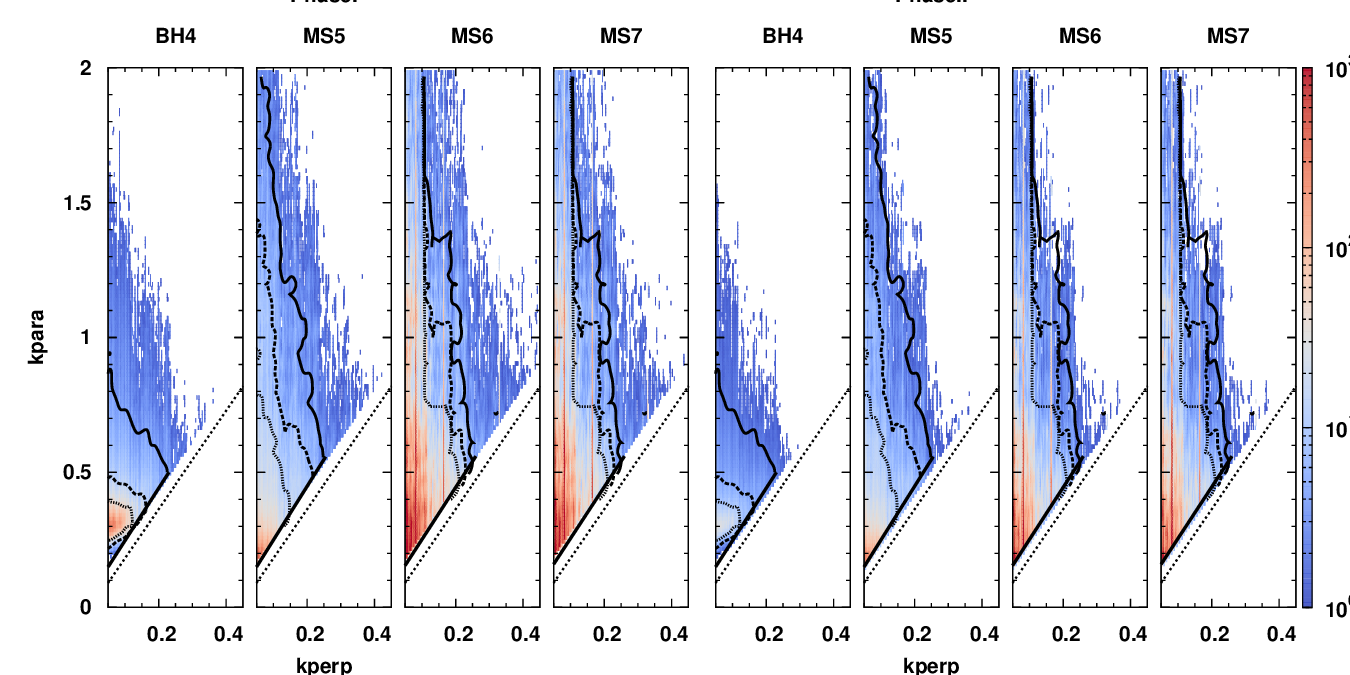}
\caption{The ratio $\R$ (eq.~\ref{eq:r}) for different window functions considered in this work. The left and right panels show the results for the Uniform and Triangular illumination respectively. $\R \ge 10, 50 \, {\rm and} \, 100$ regions for the Triangular illumination are shown by the solid, dashed and fine-dotted contours respectively in both the left and right panels. The Triangular illumination considered here represents the worst possible scenario for the illumination pattern and the allowed $(k_{\perp}, \kpar)$ range for the actual OWFA beam pattern is not expected to be smaller than that of shown by the contours.}
\label{fig:FG_ratio}
\end{figure*}

To identify the $(k_{\perp}, \kpar)$ modes which can be used for the \HI power spectrum detection we 
introduce the ratio 
\begin{equation}
    \R= P_{\mathrm{T}}(k_{\perp},k_{\parallel}) / P_{\mathrm{L}}(k_{\perp},k_{\parallel})\, ,
\label{eq:r}
\end{equation}
where $P_{\mathrm{T}}(k_{\perp},k_{\parallel})$ is the theoretically expected \HI 21-cm signal power spectrum
(eq.~\ref{eq:HI_PS}) and $ P_{\mathrm{L}}(k_{\perp},k_{\parallel})$ is the foreground leakage contribution.
 Figure~\ref{fig:FG_ratio} shows $\R$ for the different higher term window functions.
The left and right panels show the results for the Uniform and the Triangular illumination respectively. 
We have only shown the points where $\R > 1$ \ie the  \HI signal exceeds the foreground leakage.
For both the illumination patterns we find that the largest  values of $\R$, which are in  the range $ 50-500$,  
are located at the lowest  $(k_{\perp}, \kpar)$ modes just beyond the wedge boundary. The values of $\R$ and the 
region where $\R>1$ both increase as we increase the number of terms in the window function. In all cases we have 
$\R<1$ at large  $(k_{\perp}, \kpar)$ where the \HI signal is small.  In comparison to the Uniform illumination, the region where $\R>1$ is found to be  somewhat smaller for the Triangular illumination because of the larger FoV.

We have assumed that the  $(k_{\perp}, \kpar)$ region where $\R \ge \Rt$ can be used to detect the \HI 21-cm signal power  spectrum.  $\Rt$ here is a threshold  value which has to be  set sufficiently high  so as to minimize the possibility of residual foreground contamination.  We discuss the criteria for deciding the value  of $\Rt$ later in this section. We see that the $(k_{\perp}, \kpar)$  region corresponding to different  values of $\Rt$
 are somewhat smaller for the Triangular illumination as compared to the Uniform illumination. The OWFA illumination pattern is unknown, but we expect the actual OWFA predictions to  be somewhere between the Uniform and the Triangular predictions.  The $(k_{\perp}, \kpar)$ range
 which simultaneously satisfies $\R \ge \Rt$ for both the Uniform and the Triangular illuminations can safely  be  used to detect the  \HI 21-cm signal power  spectrum.
 The $\R \ge \Rt$ regions for $\Rt = 10, 50 \, {\rm and} \, 100$ for the Triangular illumination are shown by the solid, dashed and fine-dotted contours respectively in both the left and right panels. 
 The Triangular illumination considered here represents the worst possible scenario for the illumination pattern of the OWFA antennas. We do not expect the allowed $(k_{\perp}, \kpar)$ range for the actual OWFA beam pattern to be smaller than that predicted for the Triangular illumination. Thus for any value of $\Rt$, throughout we have used the Triangular illumination  to determine the allowed  $(k_{\perp}, \kpar)$ range.  

From Figure~\ref{fig:FG_ratio} we see that the allowed $(k_{\perp}, \kpar)$ region and the peak $\R$ values increase as we 
increase the number of terms  in the  window function. It thus appears to be advantageous for  \HI  21-cm signal 
detection to increase the number of  terms in the  window function. This  would indeed be true if the power spectrum estimated at 
the different $(k_{\perp}, \kpar)$ modes were uncorrelated. However, the convolution in  eq.~(\ref{eq:conv}) causes the \HI signal at different $\kpar$ modes to be correlated. We see that $\tilde{f}(\tau_m)$  gets
 wider  (right panel of Figure~\ref{fig:w_func}) causing  the $\kpar$ extent of the  correlations  to  increase as we increase the number of terms in the  window function. The system noise contribution at the different $\kpar$ modes
are also expected to be correlated  because of the convolution. Further,  the window function $F(\nu)$ gets narrower  (left panel of Figure~\ref{fig:w_func})  and   the loss in  the \HI signal   at the edge of the frequency band also  increases  as we increase the number  of terms.  It is therefore not obvious whether  it is  advantageous for  \HI  21-cm signal 
detection to increase the number of  terms in the  window function.  Rather, it would be more appropriate  to ask as to which of the different window functions considered here is best suited for \HI signal detection. 
In order to quantitatively address this issue  we consider a figure of merit namely  the Signal to Noise Ratio (SNR) for measuring $A_{\HI} = b^2_{\HI} \, \bar{x}^2_{\HI}$ which is the amplitude of the \HI 21-cm signal power spectrum (eq.~\ref{eq:HI_PS}). We have used the Fisher-matrix formalism where 
the SNR  for the measurement of $A_{\HI}$ is given by 
\begin{equation}
{\rm SNR}^2 =  \sum_{a, m, m'} \frac{\partial
    P_{\mathrm{T}}(k_{\perp a}, \kparm )}{\partial \ln A_{{\rm \HI}}} \, \mathbfit{C}^{-1}_a(m, m')\frac{\partial P_{\mathrm{T}}(k_{\perp a},k_{\parallel m'})}{\partial \ln A_{{\rm \HI}}} \, .
\label{eq:snr}
\end{equation}

Here we have assumed that  the entire allowed $(k_{\perp}, k_{\parallel})$ range where $\R \ge \Rt$ is combined to estimate $A_{\HI}$. 
Considering  $\Delta \hat{P}\left( \U_a, \tau_m \right)= \hat{P}\left( \U_a, \tau_m \right) -
{P}\left( \U_a, \tau_m \right)$ the error in estimated the 21-cm power spectrum, the correlation between different $\kpar$ mode arising from the convolution in eq.~(\ref{eq:conv})  can be quantified through the covariance matrix 
\begin{equation}
\mathbfit{C}_a(m, m')=\langle [\Delta \hat{P}\left( \U_a, \tau_m \right)] [\Delta \hat{P} \left( \U_a, \tau_{m'} \right)]
\rangle \,.
\label{eq:cov_mat}   
\end{equation}
We have estimated ${P}\left( \U_a, \tau_m \right)$ and $\mathbfit{C}_a(m, m')$   for  different window functions 
and different values of $\tobs$ using simulations. For each case  we have  used  $N_r = 1,000$  statistically independent realizations of the OWFA visibilities incorporating the   \HI signal and the system noise.

\begin{figure*}
\psfrag{BH4}{ BH4}
\psfrag{BH5}{  MS5}
\psfrag{BH6}{ MS6}
\psfrag{BH7}{ MS7} 
\psfrag{Uniform}{Uniform}
\psfrag{Triangular}{Triangular}
\psfrag{PhaseI}{ }
\psfrag{ratio10}{$\Rt = 10$}
\psfrag{ratio100}{$\Rt = 100$}
\psfrag{ratio500}{$\Rt = 500$}
\psfrag{total}{ \tiny \kern-0.3em  Total}
\psfrag{snr}{ SNR}
\psfrag{tobs}{ $\tobs$ (hours)}
\psfrag{ratio}{\qquad \qquad \qquad $\Rt$}
\psfrag{}{}
\centering
\includegraphics[scale=0.9]{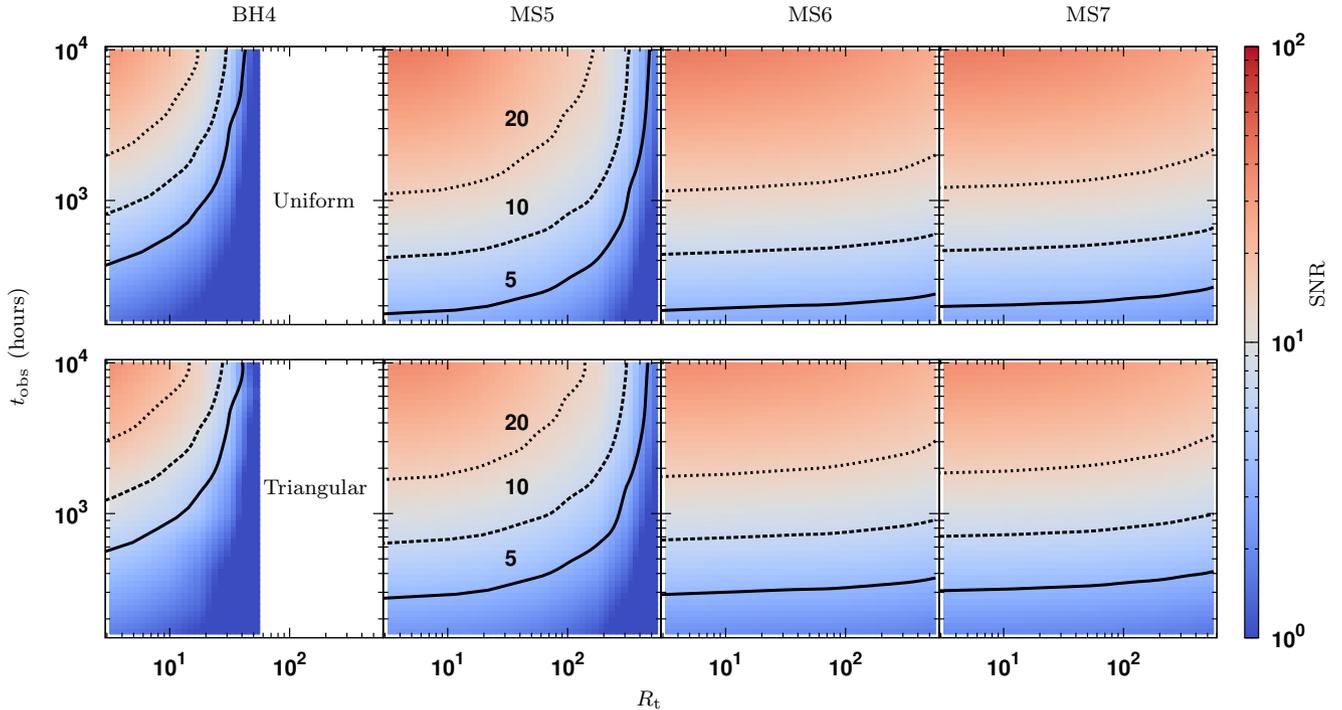}
\caption{A comparison of predicted SNRs for different higher term window functions 
considered in this work. The upper and lower panels show the predictions 
for the Uniform and Triangular illuminations respectively. The 
SNR values $5, 10 \, {\rm and} \, 20$ are shown by the solid, dashed and dotted contours respectively.}
\label{fig:snr}
\end{figure*}
Figure~\ref{fig:snr} shows the predicted SNR values as a function of $\tobs$  
and $\Rt$. The four columns respectively  correspond to the four higher term window functions, 
whereas the two rows respectively  correspond to the Uniform and Triangular illuminations. Our aim here 
is to identify the optimal window function. Considering BH4, we find that  the SNR values are  considerably 
lower  compared to  the three other window functions and BH4 is not a good choice. 
We find that  for the entire  $\Rt$  range considered here $(1 \le \Rt \le 500)$  the SNR values do not differ 
much between the MS6 and MS7  window functions. The SNR values for the MS5 window function also  are comparable to those 
for MS6 and MS7 for $\Rt \lesssim 30$,  however  the SNR values for MS5 drop rapidly 
for larger  $\Rt (> 30)$. Figure~\ref{fig:snr} therefore indicates that BH4 can definitely be excluded,  
however  all three MS5, MS6 and MS7 exhibit comparable performance if one wishes to use a threshold $\Rt < 30$.  
For a higher threshold $\Rt > 30$ MS5 also is excluded, however both MS6 and MS7 exhibit comparable performance.  
 
\begin{figure*}
\psfrag{BH4}{\small \kern-0.9em  BH4}
\psfrag{BH6}{\small \kern-0.9em  $\mathcal{R}_{\rm MS6}$}
\psfrag{BH5}{\small \kern-0.9em  MS5}
\psfrag{BH7}{\small \kern-0.9em  MS7} 
\psfrag{Uniform}{Uniform}
\psfrag{Triangular}{Triangular}
\psfrag{PhaseI}{ }
\psfrag{ratio10}{$\Rt = 10$}
\psfrag{ratio100}{$\Rt = 100$}
\psfrag{ratio50}{$\Rt = 50$}
\psfrag{total}{ \tiny \kern-0.3em  Total}
\psfrag{SNR}{\qquad \qquad \quad $\cR$}
\psfrag{obshr}{\kern-20pt $\tobs$ (hours)}
\psfrag{}{}
\centering
\includegraphics[scale=1.17, trim = 0cm 0cm 0.5cm 0cm, clip=true]{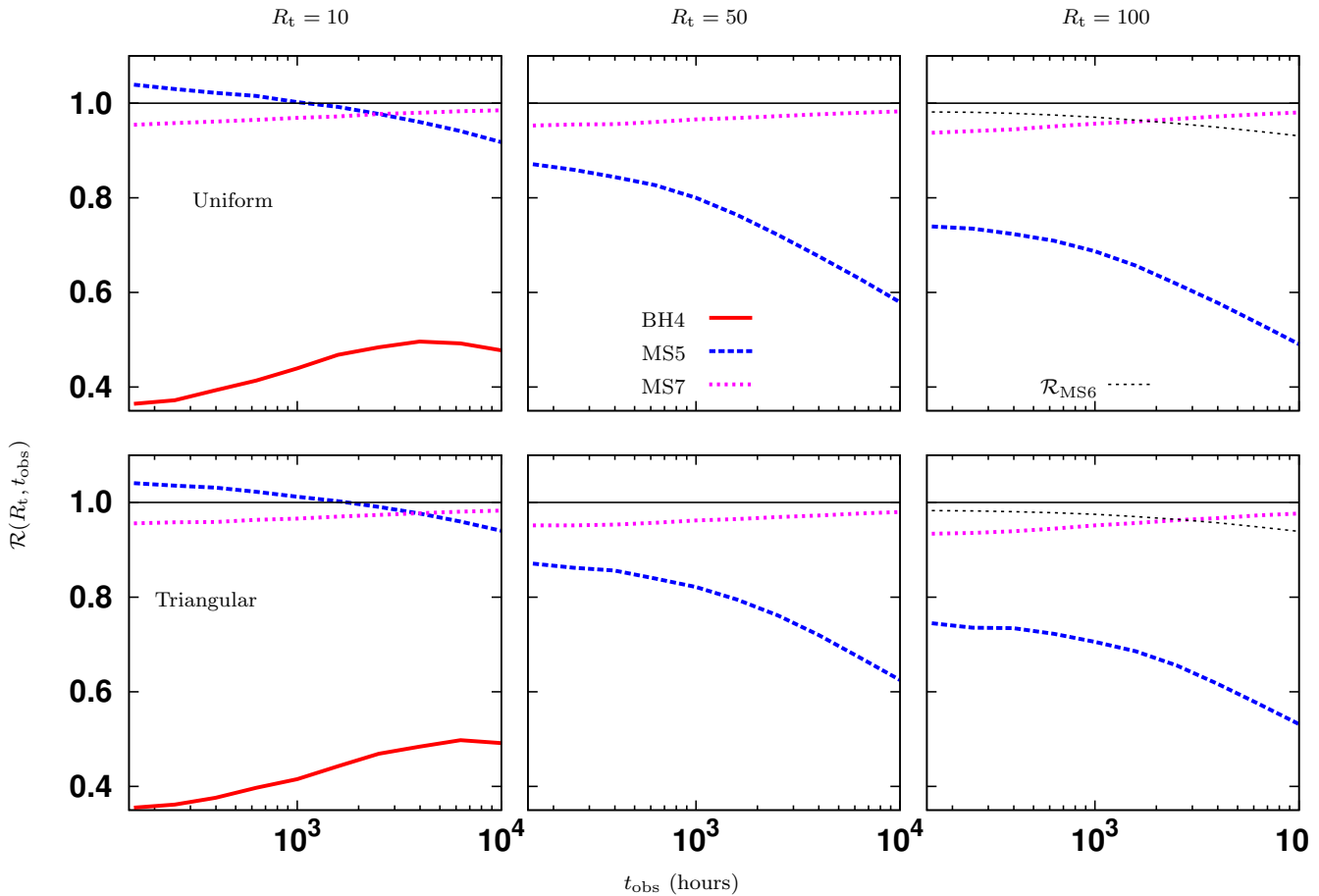}
\caption{A comparison of $\cR $  
for different window functions considered in this work. 
The upper and lower panels show the predictions for the Uniform and Triangular 
illuminations respectively. The left, middle and right panels show the results for the 
three cases where $\Rt = 10, \, 50 \, {\rm and} \, 100 $ respectively. The thin dashed line 
in the right panels show the ratio $\mathcal{R}_{\rm MS6} = [{\rm SNR}(\Rt = 100, \tobs)]_{\rm MS6} / 
[{\rm SNR}(\Rt = 50, \tobs)]_{\rm MS6}$.}
\label{fig:snr_ratio}
\end{figure*}

In order to quantify the small differences in the SNR predictions of the window functions, we consider 
the ratio of the SNRs for the different window functions with respect to that for MS6 which we take as reference
\begin{equation}
  \cR = {\rm SNR}(\Rt, \tobs)/ [{\rm SNR}(\Rt, \tobs)]_{\rm MS6} \, .  
  \label{eq:snr_ratio} 
\end{equation}
A value $\cR >1$ tells us that the corresponding window function performs better than MS6 whereas the 
converse is true if $\cR <1$. The left, middle and right panels of  Figure~\ref{fig:snr_ratio} 
show $\cR$ as a function of $\tobs$ for $\Rt = 10, 50 \, {\rm and} \, 
100$ respectively. As expected, the  $\cR$ values always remain substantially  below $1.0$ for  BH4  and this is excluded. 
Note that the $\cR$ values for BH4 are  not visible in the middle and right panels due to the very small allowed 
$(k_{\perp}, k_{\parallel})$ region at these $\Rt$ values.
Considering MS5 next, for  $\Rt = 10$ we find that  
$\cR \geq 1.0 $  provided $\tobs \leq 1,000$ hours, however  $\cR <1$ if 
$\tobs > 1,000$ hours and it declines steadily with increasing $\tobs$. 
For $\Rt = 50 \, {\rm and} \, 100$,  we have $\cR <1$ irrespective of $\tobs$. 
Considering MS7, we find that $0.9< \cR < 1.0$ for all the three $\Rt$ values shown here. 
This is a direct consequence of the fact that the  extent of the  correlation between the $\kpar$ modes increases (Figure~\ref{fig:w_func})  with an increase in the  number of terms in the window function. Although the allowed  $(k_{\perp}, k_{\parallel})$ region  increases if we increase the number of terms, the enhanced correlation causes the SNR to degrade beyond MS6. 
The Uniform and Triangular illuminations both show very similar results. 
Our  analysis suggests that the MS5  window is optimal at small $\Rt$ (\eg $\Rt \leq 30$) and small $\tobs$ (\eg $\tobs \leq 1,000$ hours),  barring this situation the   MS6 window function is optimal for  \HI power spectrum estimation with OWFA. 

Once we have identified the optimal window function,  we next aim to fix a suitable $\Rt$ for  
\HI 21-cm power spectrum estimation. We have earlier discussed that the value of $\Rt$ must be set sufficiently high to minimize the possibility of residual foreground contamination. Shorter observations (\eg $\tobs \leq 1000$ hours) are expected to have a relatively large noise contribution,  and it is possibly adequate to  consider a less conservative threshold $\Rt \approx 10$ along with the MS5 window function for \HI power spectrum estimation. For $\tobs \ge 1000$ hours where we target a more precise measurement  of the \HI power spectrum,  it is worth considering a more conservative  threshold $\Rt \ge 50$ and use the MS6 window function. The question is whether the ${\rm SNR}$ would fall significantly if we increase the value of the threshold $\Rt$ in the range $50$ to $100$. 
The thin dashed line 
in the right panels of Figure~\ref{fig:snr_ratio} show the ratio $\mathcal{R}_{\rm MS6} = [{\rm SNR}(\Rt = 100, \tobs)]_{\rm MS6} / 
[{\rm SNR}(\Rt = 50, \tobs)]_{\rm MS6}$. We find that  the SNR values degrade at most 
by $\sim 8\%$ if we increase  $\Rt$ from $50$ to $100$. This indicates that one can set the value  of the threshold $\Rt$ as high as 
$100$ without a significant loss of SNR. For $\Rt = 100$, the residual foreground contamination is  expected to be $\le  1 \%$  for every $(k_{\perp}, k_{\parallel})$ modes that is used for \HI power spectrum estimation .

\begin{figure}
\psfrag{Uniform}{ Uniform}
\psfrag{Triangular}{ Triangular}
\psfrag{PhaseI}{PII}
\psfrag{R10}{$\Rt = 10$}
\psfrag{R100}{$\Rt = 100$}
\psfrag{SNR}{ SNR}
\psfrag{tobs}{ $\tobs$ (hours)}
\psfrag{loss}{\kern-0.5em $\Delta{\rm SNR}\,({\rm in} \%)$}
\psfrag{k}{ \kern-0.5em $k \, {\rm Mpc}^{-1}$}
\psfrag{1000}{ \small \kern-1.30em  $10^3\,{\rm Hrs}$}
\psfrag{4000h}{ \small \kern-2.5em  $4\times10^3\,{\rm Hrs}$}
\psfrag{10000h}{ \small \kern-0.4em  $10^4\,{\rm Hrs}$}
\psfrag{SNRcb}{SNR}
\centering
\includegraphics[scale=0.75]{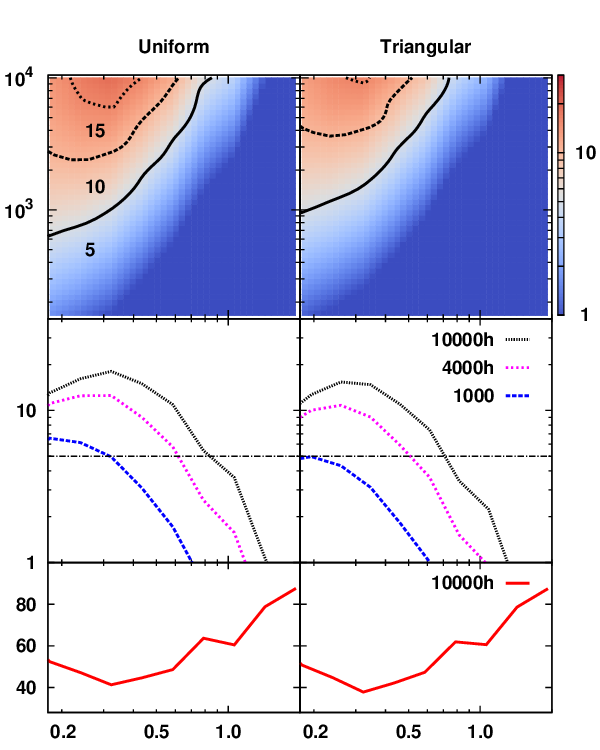}
\caption{The SNR predictions for the binned power spectrum estimation using MS6 window function 
for the Triangular and Uniform illuminations. Here we have set $\Rt = 100$.
The upper panels show the predicted SNR as a function of $k$ and $\tobs$. 
The contours mark the SNR values $5, 10 \, {\rm and} \, 15$ (mentioned in the figure).
The middle panels show horizontal sections through the upper panels for $\tobs = 1000, 4000 \, {\rm and} \, 
10^4$ hours (mentioned in the figure legend). The horizontal dot-dashed line marks the SNR value 5. The lower panels  show the percentage loss of SNR $(\Delta{\rm SNR})$ 
due to the presence of the foregrounds for $\tobs = 10^4$ hours.}
\label{fig:bin_snr}
\end{figure}
An earlier study \citep{Sarkar2016a} has predicted that a $5 \, \sigma$ detection of the binned power spectrum is possible in the $k=\sqrt{k^2_{\perp}+ k^2_{\parallel}}$ range 
$0.05 \le k \le 0.3 \, {\rm Mpc}^{-1}$ with $1,000$  hours of observation, this however 
uses the  entire available $(k_{\perp}, k_{\parallel})$  region and 
does not take the foreground contamination  into account.  The  fact is that a significant $(k_{\perp}, k_{\parallel})$ range has to be  excluded due to the foreground wedge and the residual foreground leakage. 
We next consider the revised SNR predictions for the binned \HI power spectrum taking into account the $(k_{\perp}, k_{\parallel})$  modes which have to be excluded to avoid the foreground contamination. 
For these prediction we have used the MS6 window function and set a high  threshold  of
$\Rt = 100$, the results do not change very much  if $\Rt$ is varied in the range  $\Rt = 10 \, {\rm and} \, 100$ (Figure~\ref{fig:snr}). The  range $k  \le 0.1 \, {\rm Mpc}^{-1}$ is completely within the foreground wedge, and this is excluded from \HI power spectrum estimation. We have 
binned the allowed $k$ range ($0.1 < k < 2.0 \, {\rm Mpc}^{-1}$) into $10$ logarithmic bins and estimated the SNR prediction for different $\tobs$.  The upper row of Figure~\ref{fig:bin_snr} show the SNR predictions  as a function of $k$  and $\tobs$,  
with the left and right panels corresponding to 
the Uniform and Triangular illuminations respectively. The middle row shows  horizontal sections through the upper panels 
\ie they show the SNR as a function of $k$  for fixed values of $\tobs$ (mentioned in the figure legend), 
and the lower row shows the percentage loss of SNR ($\Delta{\rm SNR}$)  due to the excluded 
$(k_{\perp}, k_{\parallel})$  region. To calculate $\Delta{\rm SNR}$   we have used the SNR predictions 
considering the entire available $(k_{\perp}, k_{\parallel})$  region (similar to \citealt{Sarkar2016a}) as reference. 

Considering the upper row of Figure~\ref{fig:bin_snr}, we see that the SNR predictions are similar for both the illuminations but the SNR values are $\sim 1.5$ times lower for the Triangular illumination in comparison to the Uniform illumination. Our results are also similar to those in  Figure 3 of \citet{Sarkar2016a} except that  our prediction for the Uniform illumination are  $\sim 1.5$ times lower due to the foreground contamination. We find that at low $\tobs$ the SNR peaks in the smallest $k$ bin $(\sim 0.18 {\rm Mpc}^{-1})$ and  a $5\,\sigma$ measurement is possible at this $k$ bin with 
$\tobs \approx  600$ hours and $1,000$ hours for the Uniform and Triangular illuminations respectively.
A $5\,\sigma$ detection of the binned power spectrum is possible in the $k$ range $0.18 \leq k \leq 0.3 \, {\rm Mpc}^{-1}$  with $\tobs \sim 1,000$ hours for the Uniform illumination,  
whereas this  will require  $\tobs \sim 2,000$ hours for the Triangular illumination.
The peak SNR shifts towards  larger $k$ bins for larger  $\tobs$,  and the  peak is at $ k \sim 0.3 \, {\rm Mpc}^{-1}$ for $\tobs = 10^4$ hours where a $15 \, \sigma$ detection is possible. The shift in the peak SNR  is clearly visible   in the middle row of the figure. 
 A $10 \, \sigma$ detection is possible  in the range $k \sim 0.2-0.4 \, {\rm Mpc}^{-1}$  with $\tobs \sim 3,000$ hours and $4,000$ hours for the Uniform and Triangular illuminations respectively. 
 The SNR falls drastically at large $k$ $(>0.8\, {\rm Mpc}^{-1})$, this is also noticeable in  
 Figure 3 of \citet{Sarkar2016a} and this is due to the fact that the \HI power spectrum fall at large $k$ (Figure \ref{fig:pk}) whereby  these bins are dominated by the system noise contribution. 
 The situation is further aggravated here because a considerable fraction of the available  
$(k_{\perp}, k_{\parallel})$  region has to be excluded to avoid the foregrounds. Considering the lower row of the figure, we see that the fractional loss in the SNR $(\Delta {\rm SNR})$ is $> 60 \%$
at $k >0.8\, {\rm Mpc}^{-1}$, and it  increases rapidly to $\sim 80\%$ at the  larger $k$ bins. 
The  fractional loss in the SNR is in the range $40 - 60 \%$  for  $k$ in the range $0.18 \le  k \le 0.8 \, {\rm Mpc}^{-1}$ where there are prospects of a detection. We also note that $(\Delta {\rm SNR})$ is minimum at  $\lesssim 40\%$ at $k\sim 0.3 \, {\rm Mpc}^{-1}$ where the SNR peaks for  $\tobs \geq 10^4$ hours.

\section{SUMMARY AND CONCLUSION}\label{sec:summary}
The ORT \citep{Swarup1971} is currently being upgraded to operate as a
radio interferometer, the Ooty Wide Field Array (OWFA;
\citealt{Subrahmanya2016b}) and this work focuses on PII of OWFA.  
The array operates with a single linear polarization. 
The ORT  (and also OWFA) feed system consists  of linear dipoles arranged end to end along the 
long axis of the cylindrical parabolic reflector. Considering any particular dipole, its 
radiation pattern is minimum along the direction of 
the adjacent dipoles and we  thus expect minimal coupling  between the adjacent dipoles.
The actual primary beam patten $\pb$ for OWFA is unknown. For 
this study we use two extreme models for $\pb$, the first one is based 
on the simplest assumption that the OWFA antenna aperture is uniformly illuminated 
by the dipole feeds (Uniform illumination) whereas the second one assumes 
a Triangular illumination pattern (Figure~\ref{fig:Pbeam}). We expect the actual OWFA 
illumination to be somewhat in between these two scenarios. 

OWFA is sensitive to the HI 21-cm signal from $z=3.35$, and measuring the cosmological 21-cm power spectrum is one of the main goals of this upcoming instrument.  The cosmological \HI 21-cm signal is faint and  is buried in  foregrounds which are several  orders of  magnitude brighter. The foregrounds processed  
through  the chromatic response of the instrument produce spectral features  which  contaminate the \HI signal, and this poses a severe challenge for detecting the 21-cm power spectrum. 
In this paper we have simulated the \HI 21-cm signal and foregrounds 
expected for OWFA PII. Our aim here is to use these simulations to quantify the extent  of the expected foreground  contamination and asses the prospects  of detecting the 21-cm power spectrum. 

We have used all sky foreground simulations (described in Section~\ref{sec:fg})  which 
 incorporate  the contributions from the  two most dominant components namely the diffuse Galactic synchrotron emission and the extragalactic point sources. These were used to calculate 
the foreground contribution  $\fgvisai$ to the model visibilities (eq.~\ref{eq:modvis1}) expected at OWFA. 
These simulations incorporate the chromatic behaviour of both the sources and also the instrument.  
To simulate the \HI signal contribution to the model visibilities $\sigvisai$ 
(eq.~\ref{eq:vis3}), we use the ``Simplified Analysis" presented in \citet{Sarkar2016b}.  
This is based on the flat-sky approximation,  and also 
ignores the correlation between the \HI signal at adjacent baselines 
and  the non-ergodic nature of the \HI  visibility signal along the frequency 
axis. To estimate the 21-cm power spectrum from the measured visibilities, we  introduce 
an estimator (eq.~\ref{eq:p_same}) which  has been  constructed so as to eliminate the noise bias and provide an unbiased estimate of the 3D power spectrum $P(k_{\perp}, 
k_{\parallel})$. We have validated this  for both the Uniform and the  Triangular illuminations  using  a large number of statistically 
independent realizations of  \HI simulations. These particular simulations 
also include the system noise, the foregrounds  however are ignored.   We find  (Figure~\ref{fig:pk}) that in the absence of foregrounds, for both the illuminations, the $k$ range $0.05-0.3 \, {\rm Mpc}^{-1}$ is 
most favourable for measuring the power spectrum with OWFA. This is consistent with the results of earlier work  \citep{Sarkar2016a}.

Considering the foregrounds,  the contamination  is primarily localized within  a wedge shaped region of the $(k_{\perp}, k_{\parallel})$ plane (Figure~\ref{fig:FG_PI}). The $k$-modes outside this ``foreground wedge" are believed  to be largely  uncontaminated  by the foregrounds. However,  there is a relatively small  fraction  of the foreground which leaks out beyond the wedge. Though small, this foreground leakage may still exceed the expected HI signal in many of the $k$-modes outside the foreground wedge. 
For signal detection 
we focus on a strategy referred to as ``foreground avoidance" where  only the  $k$-modes which are expected to be uncontaminated are used for measuring the 21-cm power spectrum. In this work we use simulations to identify the region of the $(k_{\perp}, k_{\parallel})$ plane which is expected to be uncontaminated, and  we use this to quantify the prospects of measuring the 21-cm power spectrum using OWFA.  

 Our simulations show that foreground leakage outside the wedge, though small, can still exceed the 21-cm power spectrum expected at OWFA. We find that the extent of foreground leakage is extremely sensitive to the frequency window function  $F(\nu)$  (eq.~\ref{eq:vis5})  which is introduced \citep{Vedantham2012}  to suppress the measured visibilities  near the boundaries  of the frequency band.  Considering  the extensively  used (e.g.  \citealt{Paul2016}) Blackman-Nuttall  filter which has four terms,   we find that the foreground leakage  exceeds the expected 21-cm power spectrum at all the available $k$ modes,  
 and it will not be possible to measure the 21-cm power spectrum using OWFA. 
In order to overcome this problem, we consider a  set of cosine window 
functions with progressively increasing number of terms (Table~\ref{tab:window} and Figure~\ref{fig:w_func}).  The window function gets narrower resulting in  better suppression at the edges of the band as we increase the number of terms. Using $\R$ which is the ratio of the expected 21-cm power spectrum to the foreground leakage contribution, we find that the 
$(k_{\perp}, \kpar)$ region where $\R > 1$ (\ie the region where \HI signal exceeds 
the foreground leakage) increases if we increase the number  of terms in the 
window function (Figure~\ref{fig:FG_ratio}). Taken at face value, this indicates that it is advantageous to increase the number of terms in the window function. It is however also necessary to take into consideration the fact that the HI signal  in adjacent  $\kpar$ modes get correlated due to $F(\nu)$  and the extent of this correlation increases as we increases the number of terms. The number of independent estimates of the 21-cm power spectrum thus gets reduced if we increase the number of terms. 
We therefore need to choose the optimal window function by balancing between these two 
competing effects.  We have used the Fisher matrix formalism to define the SNR (eq.~\ref{eq:snr}) 
for measuring the amplitude of the 21-cm  power spectrum, and we use this as a figure of merit
to identify the optimal window function. 

Our  analysis (Figure~\ref{fig:snr}) shows that the optimal choice of window function depends on the observing time $\tobs$ and the threshold value $\Rt$. A threshold value $\Rt$ implies that we only use the modes where $\R \ge \Rt$ for measuring the 21-cm power spectrum.  We note that the value of $\Rt$ must be set sufficiently high to minimize the possibility of residual foreground contamination.
We find that the five term  MS5  window function is optimal at  small $\tobs$ ($\leq 1,000$ hours) and small $\Rt$ ($\leq 30$),  whereas the six term MS6 window function is optimal for larger values of $\tobs$ and $\Rt$.   Relative to MS6, the SNR is found to degrade slightly  if we consider the  seven term MS7 window function.  The Uniform and Triangular illuminations both show very similar results. 
 
 We propose a possible observational  strategy based on the finding summarized above. 
Shorter observations (\eg $\tobs \leq 1,000$ hours) are expected to have a relatively large noise contribution,  and it is possibly adequate to  consider a relatively low threshold $\Rt \approx 10$ along with the MS5 window function.  For longer observations $\tobs \ge 1,000$ hours where we target a more precise measurement  of the 21-cm  power spectrum,  it is worth considering a more conservative  threshold $\Rt \ge 50$ and use the MS6 window function. Our investigations also show that the SNR does not fall much if $\Rt$ is increased from $50$ to $100$, and we could equally well consider using a very conservative threshold of $\Rt=100$ where the contribution from foreground leakage is expected to be 
less than $1 \% $ of the 21-cm power spectrum.   

The SNR values  for measuring the amplitude of the 21-cm power spectrum (Figure~\ref{fig:snr}) are approximately $1.5$ times lower for the Triangular illumination in comparison to the Uniform Illumination.  Using  MS5 with $\Rt  \approx  30$,  a $5\sigma$ detection will take $\sim 180$ hours and $\sim 300$ hours  with the Uniform   and Triangular illuminations respectively. The same is increased to 
$\sim 200$ hours and $\sim 300$ hours  if we use MS6 or MS7 with $\Rt \approx 100$.

We have also considered the prospects of measuring the binned 21-cm power spectrum. The discussion here is restricted to MS6 with $\Rt=100$. We find that the   range $k  \le 0.1 \, {\rm Mpc}^{-1}$ is completely within the foreground wedge (Figure~\ref{fig:bin_snr}) and has to be excluded. For low  $\tobs$ the SNR peaks at the smallest 
$k \approx 0.18 {\rm Mpc}^{-1}$ bin and  a $5\,\sigma$ measurement is possible at 
this $k$ bin with $\tobs \approx  600$ hours and $1000$ hours for the Uniform and 
Triangular illuminations respectively. A $5\,\sigma$ detection of the binned power spectrum 
is possible in the $k$ range $0.18 \leq k \leq 0.3 \, {\rm Mpc}^{-1}$  with $\tobs \sim 1,000$ 
hours for the Uniform illumination, whereas this  will require  $\tobs \sim 2,000$ hours for 
the Triangular illumination. Considering $\tobs=10^4$ hours, for both the illuminations the peak SNR shifts to larger $k$ values $ 0.3 -0.4 \,  {\rm Mpc}^{-1}$ and a $5\,\sigma$ detection is possible in the range $0.18 \le  k \le 0.8 \, {\rm Mpc}^{-1}$.   We have used $\Delta {\rm SNR}$ to quantify the fractional loss in 
SNR due to the foreground contamination, the comparison here is with respect to the situation where there are no foregrounds.  We find that  $\Delta {\rm SNR}$  has values  in the range $40 - 60 \%$  for  $k$ in the range $0.18 \le  k \le 0.8 \, {\rm Mpc}^{-1}$ where there are good prospects of measuring the 21-cm power spectrum. 

The exact beam pattern of OWFA is not known, but we expect this to be somewhere between the Uniform and Triangular illuminations considered here. We therefore expect the actual situation for measuring the 21-cm power spectrum to lie somewhere between the two different sets of predictions presented here.  
The present study indicates that ``Foreground Avoidance" provides an effective technique for measuring the 21-cm power spectrum with OWFA.  It is also predicted that a  $5 \, \sigma$ measurement of the 21-cm power spectrum should be possible within approximately a few hundred hours of observations despite the $k$ modes which have to be excluded due to foreground contamination.   It is however  necessary to note  that the entire analysis presented here is based on $20$ statistically independent realizations of our specific foreground model. While this model attempts to  incorporate the salient features of the two dominant foreground components, it still remains to establish how robust the results are with respect to variations in the foreground model. Although the exact quantum of foreground leakage may vary depending on the foreground model, we do not expect this to be a very severe effect as we have adopted a pretty conservative threshold $\Rt=100$ for a considerable part of our analysis.  
Calibration \citep{Marthi2014} is another  issue which could affect the results presented here. In future work we plan to study the effect of calibration errors and also the effect of varying the foreground model.


\section*{ACKNOWLEDGEMENT}
The authors would like to thank Jayaram N Chengalur for useful suggestions and discussions. 
SC acknowledges the University Grant Commission, India for
providing  financial support through Senior Research Fellowship. SC would also like to thank 
Abinash Kumar Shaw for helpful discussions and suggestions. VRM acknowledges support of the Department of Atomic Energy, Government of
India, under project no. 12-R\&D-TFR-5.02-0700.

\section*{DATA AVAILABILITY} 
The data underlying this article will be shared on reasonable request to the corresponding author.

\bibliographystyle{mnras} 
\bibliography{mylist}
\vspace{1cm}
\appendix{\bf{APPENDIX}}
\section{Variance of the estimator} \label{sec:appendix}
In Section~\ref{sec:3DPS} we have used several statistically independent realizations of 
the signal to determine the variance $(\sigma^2)$ of the estimated power spectrum. 
Such a procedure is, by and large, only possible with simulated data. We
usually have accessed to only one statistically independent realizations of the sky  signal, and the aim is to use this to not
only estimate   the power spectrum but also predict  the uncertainty in the estimated power spectrum. Considering the power spectrum estimator $\hat{P}\left( \U_a, \tau_m \right)$, 
we  theoretically  calculate  the variance 
\begin{equation}
\sigma_{\mathrm{p}}^2 (\U_a, \tau_m)= \langle \hat{P}^{2}\left( \U_a, \tau_m \right) \rangle - \langle \hat{P}\left( \U_a, \tau_m \right) \rangle^2 \,.
\label{eq:def_var}
\end{equation}
which is used to predict  the uncertainty in the estimated power spectrum $P(\ppk, \kpar)$.
The entire analysis here is based on the assumption that the \HI signal is a Gaussian random field.  

Considering the power spectrum estimator we have 

%
\begin{equation}
\sigma^2_p (\U_a, \tau_m) = \left[P(\ppka, \kparm) +  P_{\mathrm{N}}(\ppka, \kparm) \right]^2 \, .
\label{eq:var_same}
\end{equation}
We use $\langle \vert \nvisait \vert^2 \rangle=2 \sigma^2_N(\U_a)$ (eq.~\ref{eq:noise}), 
eq.~(\ref{eq:cf}) and (\ref{eq:pn}) to simplify the noise power spectrum,
\begin{equation}
P_{\mathrm{N}}(\ppka, \kparm) = \frac {r^2 \, r' \, T_{\mathrm{sys}}^2}{\tilde{\eta} \, \Delta t \, \Ns \, (N_{\mathrm{A}} - a) } \, ,
\label{eq:apn-np}
\end{equation}
where $\tilde{\eta} = [\int A^2(\theta) d^2\theta]/[\int A(\theta) d^2\theta]^2$ 
is a dimensionless factor  \citep{Chatterjee2018b}.
It is worth noting that the second term in the rigth-hand side of eq.~(\ref{eq:p_same}) which has been introduced 
to subtract out the noise bias in eq.~(\ref{eq:v2_same}) is ignored for calculating the variance.  The signal 
contribution from this term to the estimator is of the order of $\sim 1/\Ns$ which is extremely small for a 
long observation. 


%

%
\begin{figure}
\psfrag{Uniform}{Uniform}
\psfrag{kk}{$k$}
\psfrag{sigp}{$\sigma_{\mathrm{p}}(k)$}
\psfrag{analytical}{Theory}
\psfrag{simulation}{Simulation}
\centering
\includegraphics[scale=0.75, angle = 0]{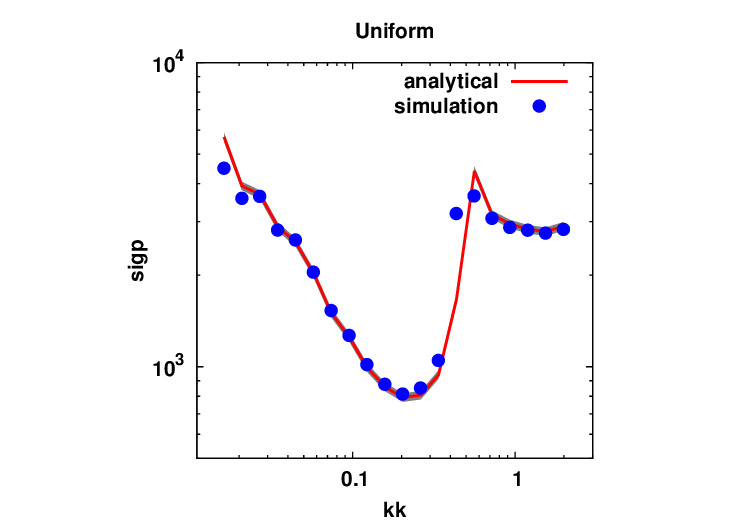}
\caption{The analytic prediction for the variance (eq.~\ref{eq:var_same}) for auto-correlation is 
compared with variance estimated from $N_r = 1,000$ realizations of the simulated signal visibilities.}
\label{fig:pk_var}
\end{figure}

Figure~\ref{fig:pk_var} shows the analytic prediction for the variance calculated using eq.(~\ref{eq:var_same}) (solid line)
for a total observation time of $\tobs = 1, 000$ hours with an integration time $\Delta t = 1$ hour for the 
Uniform illumination. For comparison we also show (points) the variance estimated  from $N_r = 1,000$ independent 
realizations of the simulated signal visibilities. Here we have binned the variance $\sigma_{\mathrm{p}}(\U_a, \tau_m)$ at the 
$(\ppk, \kpar)$ modes corresponding to the OWFA baselines and delay channels into 20 equally spaced logarithmic bins 
to compute $\sigma_{\mathrm{p}}(k)$. The shaded region in the figure shows the theoretically estimated error 
$\Delta_{\sigma_{\mathrm{p}}} = \sigma_{\mathrm{p}}(k)/ \sqrt{N_r}$ in $\sigma_{\mathrm{p}}(k)$ for $N_r = 1000$ statistically independent realizations 
of the \HI signal. We see that the analytic predictions are in reasonably good agreement
with the values obtained from the simulations over the entire $k$-range that we have considered here, except 
the two smallest $k$ bins. This discrepancy possibly arises because the estimator ignores the 
convolution with the aperture power pattern which is included in the simulated visibility signal (eq.~\ref{eq:viscor}). 
From Figure~\ref{fig:pk_var} we also notice that $\sigma_{\mathrm{p}}(k)$ remains relatively 
small in the $k$-range $0.05 - 0.3 \, {\rm Mpc}^{-1}$, which is consistent with the findings of 
\citet{Sarkar2016a}. The $k$-modes larger than $0.7 \, {\rm Mpc}^{-1}$ remains 
noise dominated and larger hours of observation is required to extract signal from these modes.

\end{document}